\documentclass[3p,times]{elsarticle}

\usepackage{xspace}
\usepackage[T1]{fontenc}
\usepackage[scaled]{beramono}
\usepackage{listings}
\usepackage{tikz}
\usepackage{dingbat}
\usepackage{amssymb}
\usepackage{amsmath}
\usepackage{bbding}
\usepackage{pbox}
\usepackage{scrextend}
\usepackage{hyperref}
\usepackage{multirow}

\def\reff@jnl#1{{\rm#1\/}}
\def\aj{\reff@jnl{AJ}}         
\def\araa{\reff@jnl{ARA\&A}}      
\def\apj{\reff@jnl{ApJ}}        
\def\apjl{\reff@jnl{ApJ}}        
\def\apjs{\reff@jnl{ApJS}}       
\def\aap{\reff@jnl{A\&A}}        
\def\aapr{\reff@jnl{A\&A~Rev.}}     
\def\aaps{\reff@jnl{A\&AS}}       
\def\mnras{\reff@jnl{MNRAS}}      
\def\physrep{\reff@jnl{Physics Reports}}
\def\prd{\reff@jnl{Phys.Rev.D}}     
\def\prl{\reff@jnl{Phys.Rev.Lett}}   
\def\pasp{\reff@jnl{PASP}}       
\def\pasj{\reff@jnl{PASJ}}       
\def\nat{\reff@jnl{Nature}}       
\def\jcap{\reff@jnl{JCAP}}   
\def\memsai{\reff@jnl{MemSAI}} 
\def\na{\reff@jnl{New Astronomy}}       

\newcommand{\hide}{\texttt{HIDE}\xspace}
\newcommand{\seek}{\texttt{SEEK}\xspace}
\newcommand{\ivy}{\texttt{Ivy}\xspace}
\newcommand{\healpix}{\texttt{Healpix}\xspace}
\newcommand{\hope}{\texttt{HOPE}\xspace}

\begin{document}

\begin{frontmatter}

\title{HIDE \& SEEK: \\ End-to-End Packages to Simulate and Process Radio Survey Data}

\author{Jo\"el Akeret\corref{cor1}}
\ead{jakeret@phys.ethz.ch}

\author{Sebastian Seehars}
\author{Chihway Chang}
\author{Christian Monstein}
\author{Adam Amara}
\author{Alexandre Refregier}

\cortext[cor1]{Corresponding author}

\address{ETH Zurich, Institute for Astronomy, Department of Physics, Wolfgang Pauli Strasse 27, 8093 Zurich, Switzerland}

\begin{abstract}

As several large single-dish radio surveys begin operation within the coming decade, a wealth of radio data will become 
available and provide a new window to the Universe. In order to 
fully exploit the potential of these data sets, it is important to understand the systematic effects associated with the 
instrument and the analysis pipeline. A common approach to tackle this is to forward-model the entire 
system -- from the hardware to the analysis of the data products.
For this purpose, we introduce two newly developed, open-source Python packages: the HI Data Emulator (\hide) 
and the Signal Extraction and Emission Kartographer (\seek) for simulating and processing single-dish radio survey data. 
\hide forward-models 
the process of collecting astronomical radio signals in a single-dish radio telescope instrument and 
outputs pixel-level time-ordered-data. \seek processes the time-ordered-data, removes 
artifacts from Radio Frequency Interference (RFI), automatically applies flux calibration, and aims to recover 
the astronomical radio signal. The two packages can be used separately or together depending on the application. 
Their modular and flexible nature allows easy adaptation to other instruments and data sets.
We describe the basic architecture of the two packages and examine in detail the noise and RFI modeling in \hide, 
as well as the implementation of gain calibration and RFI mitigation in \seek. 
We then apply \hide \& \seek to forward-model a Galactic survey in the frequency range 990 -- 1260 MHz 
based on data taken at the Bleien 
Observatory. For this survey, we expect to cover 70\% of the full sky and achieve a
median signal-to-noise ratio of 
approximately 5 -- 6 in the cleanest channels including systematic uncertainties. However, we also point out the
potential challenges of high RFI contamination and baseline removal when examining the early data from the Bleien Observatory. 
The fully documented \hide \& \seek packages are available at 
\url{http://hideseek.phys.ethz.ch/} 
and are published under the GPLv3 license on GitHub.

\end{abstract}

\begin{keyword}

Radio cosmology \sep Forward modeling \sep RFI mitigation \sep HIDE \sep SEEK

\end{keyword}

\end{frontmatter}

\section{Introduction}
\label{sec:introduction}

Forward-modeling has become a common approach in various fields of astronomy where mock 
data sets are simulated and analyzed in parallel with the science data. This has become especially 
prevalent in cosmology where large data sets are used and high precision is required. Prominent 
examples are analyses of the cosmic microwave background \citep{reinecke2006simulation}, spectroscopy 
\citep{nord2016spokes} and weak gravitational lensing \cite[e.g.][]{bridle2009handbook, 
refregier2014way, bruderer2015calibrated, peterson2015simulation}. These forward-modeling pipelines 
simulate the astrophysical signals, the instrument response and the data reduction process in order to 
understand any systematic biases from hardware or software and to estimate statistical errors in the 
measurement chain. 

In this paper, we implement this forward-modeling approach for single-dish radio surveys. Several 
single-dish radio surveys are being planned for the next decades with the goal of mapping the H$_{\rm I}$ 
neutral hydrogen in the Universe \citep{Battye2013, Santos2015, Bigot-Sazy2016}.
We develop two software packages: the {\texttt{HI Data Emulator}\xspace} (\hide) and the {\texttt{Signal 
Extraction and Emission Kartographer}\xspace} (\seek). \hide forward models the entire radio survey system 
chain, while \seek processes both the simulated data and the observed survey data in a reproducible and 
consistent way. Various sophisticated simulation and data reduction pipeline packages for radio astronomy 
exist \citep[e.g.,][]{swinbank2015lofar, mcmullin2007casa, dodson2016imaging}. However, 
many of them are either non-open source or project-specific. 
\hide \& \seek are developed in a different angle -- the initial functionalities are rather simple, but can be 
expanded easily as the codes are designed with 
a high level of modularity, flexibility and transparency, in a pure Python implementation with rigorous testing.
Developing the two packages simultaneously has the advantage that the individual 
components of one pipeline can be cross validated against its counter part in the other pipeline.

\hide \& \seek are developed based on the hardware system and data products from the 7m telescope 
at the Bleien Observatory as described in \cite[][hereafter C16]{chang2016an}. This 
framework is then used to forward-model a Galactic survey in the frequency range 990 -- 1260 MHz 
conducted at the Bleien Observatory for testing and 
science verification purposes. Such an analysis allows us to forecast the expected power of this survey
with the existing hardware system at Bleien. Comparing the results of the forward model and data also helps 
to identify areas that require improvements in \hide \& \seek as well as the hardware system. 

This paper is organized as follows. In Section \ref{sec:pipelines} we first describe the basic architecture and 
design of \hide \& \seek. Detailed implementations for specific functionalities are described in 
\ref{sec:beam_convolution}, \ref{sec:rfi} and \ref{sec:flux_calibration}. 
In Section \ref{sec:application2bgs}, we apply \hide \& \seek to forward-model a survey based 
on early data taken at the Bleien Observatory. This includes customizing the various functionalities to this specific survey 
and providing a forecast for the expected outcome of the survey. Finally, we conclude in Sections \ref{sec:conclusion}. 
In \ref{sec:comparison} we show an example of how we applied \seek to process part of the early data from 
the Bleien Observatory and what we learn comparing these results to the \hide simulations. 
Information for downloading and installing \hide \& \seek, as well as the default 
file format is described in \ref{sec:distribution} and \ref{sec:fileformat}, respectively.

\section{The HIDE \& SEEK pipelines}
\label{sec:pipelines}

\begin{figure}[t]
\begin{center}
\vspace{-0.4in}
\includegraphics[width=0.8\linewidth]{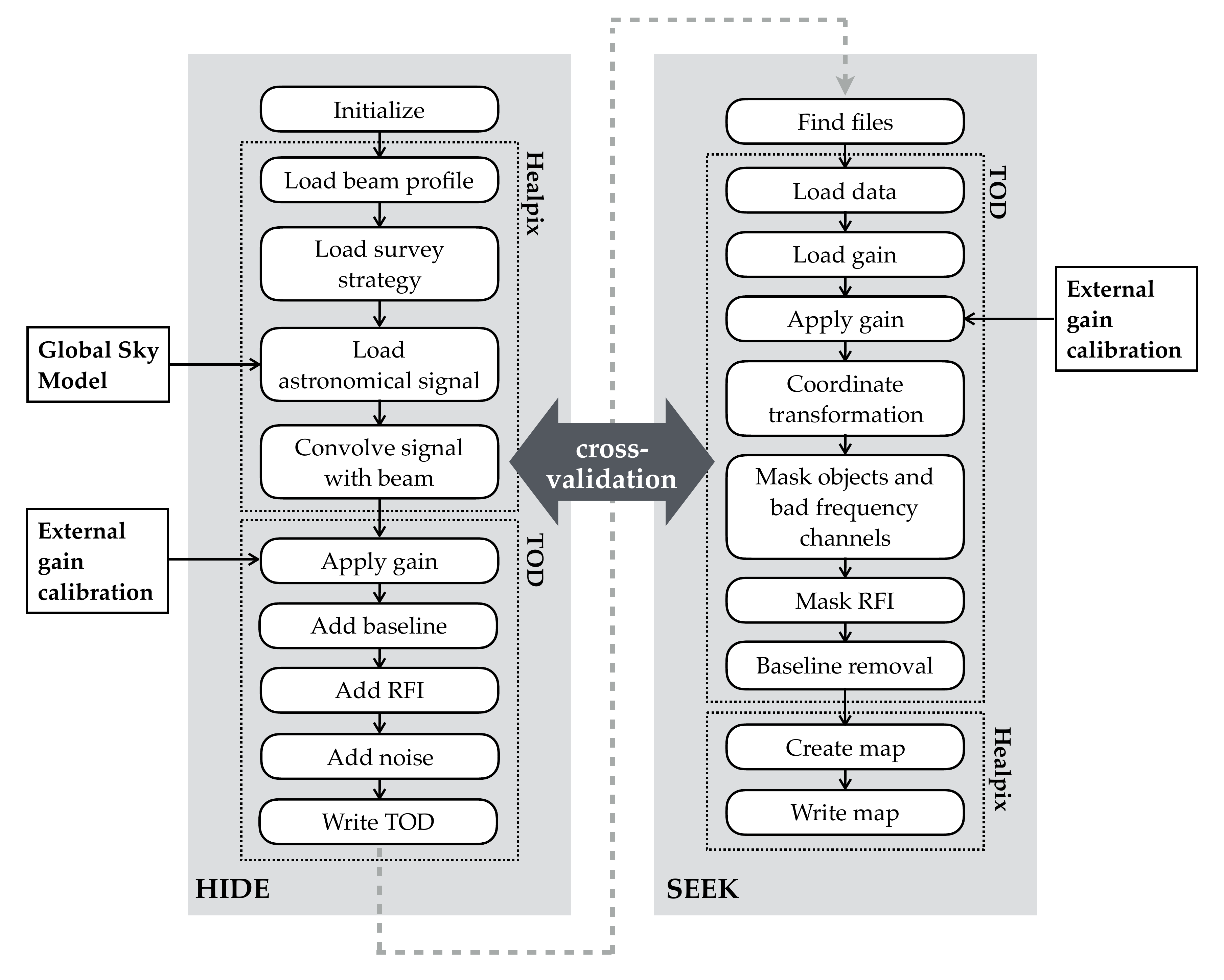}
\end{center}
\caption{Flow diagram of the steps executed in \hide (left) and \seek (right). Each box with rounded 
corners is a plugin and the other boxes indicate external data. The dotted boxes indicate the data 
format (\healpix map or TOD) along the process. For \hide, the simulations start on a \healpix map 
and outputs in TOD format. This TOD can 
be directly fed into \seek, where it aims to reconstruct the TOD back onto a \healpix map. The two 
packages together close the end-to-end loop of this forward-modeling framework. This parallel 
simulation and analysis setting also facilitates cross-validation during the development of the 
packages. }
\label{fig:flow}
\end{figure}

\hide is a simulation pipeline for single-dish radio telescopes and \seek is a data processing pipeline for observed or 
simulated radio telescope data. We have developed the two independent software packages simultaneously, which 
means that one pipeline can be used to cross validate the other. For example, in \hide we simulate radio frequency 
interference (RFI) signals, while in \seek we detect and mask the RFI signals. This suggests that the quality 
of the RFI masking in \seek can be assessed using simulated data from \hide. On the other hand, the 
goodness of the modeling in \hide can be verified by processing real data with \seek and comparing the results 
with the simulation. Both pipelines share a common design. Fig. \ref{fig:flow} shows a schematic 
illustration of both packages. 

To ensure the common design and facilitate the switching between \hide \& \seek, both package 
are based on the simple plugin-based workflow engine \ivy, which we introduce in Section \ref{sec:ivy}. 
In Section \ref{sec:data} we briefly describe the two main data structures that are used in \hide \& \seek. 

\subsection{Ivy: Plugin-based workflow engine}
\label{sec:ivy}

\ivy is a generic workflow engine written in Python and is open-source under the GPLv3 
license\footnote{\url{http://github.com/cosmo-ethz/ivy}}. Its architecture ensures that individual 
elements of pipelines (which we refer as ``plugins'' here) are self-contained. 
The interaction between the plugins is done via a context object that is passed from plugin to 
plugin by the framework. \ivy's design fosters reusability and maintainability of the code. As every 
plugin is loosely coupled to the pipeline and the other plugins, unit testing the individual components 
becomes greatly simplified. Furthermore, new features can easily be added to an existing pipeline in 
the form of a new plugin without interfering with the other plugins.

A pipeline developed with the \ivy framework always consists of a configuration file, that contains a 
list of the plugins that belong to this pipeline and parameters that are used by the plugins.
An \ivy pipeline can be executed in parallel on multiple CPU cores with minimal work. By default \ivy 
uses Python's built-in {\texttt{multiprocessing}\xspace} package but alternative parallelization scheme 
such as {\texttt{IPython cluster}\xspace} can be chosen. The workload is automatically distributed 
among the available CPU's and \ivy executes the plugins in parallel.

\subsection{Data structures: time-ordered data and \healpix maps}
\label{sec:data}

The two relevant data structures used in \hide \& \seek are time-ordered data (TOD) and \healpix 
\citep{Gorski2005} maps. TOD refers to the data type we record at the end of the instrument chain 
(a bolometer or a spectrometer). Typically, one or multiple values that scale linearly with the signal 
received by the telescope is recorded over time. In our case, since the measurement instrument 
is a spectrometer, we have one value per frequency channel per time, which results in a 2D plane with 
time and frequency on the two axes (see Fig. \ref{fig:tod} for an example of the observed and simulated TOD). 
This data format itself is agnostic about where the signal is from in 
the sky. Only by combining the data with the telescope pointing we map the TOD to the sky coordinates.

A Hierarchical Equal Area isoLatitude Pixelation (\healpix) map refers to a specific pixelation scheme 
implemented on a sphere. This format is commonly used in cosmology. \healpix maps have by 
construction equal-area pixels and the pixelation introduces minimal distortions and errors 
compared to other projection methods. \healpix maps can be manipulated with the Python wrapper 
\texttt{Healpy}\footnote{\url{https://healpy.readthedocs.io}}. In this work, we use \healpix maps whenever the 
celestial coordinate is relevant. This includes the input Milky Way signal in \hide and the final 
reconstructed map in \seek.
     
\subsection{\hide architecture}
\label{sec:hidearchitecture}

\hide \& \seek both follow the plugin design concept of the \ivy framework described above. The architecture 
allows the user to easily add new features or replace existing functionality. In the following, we give a 
high-level overview of the functionalities in both packages.

\hide is a package for simulating a single dish radio telescope survey. As such, it takes \healpix maps as inputs 
and processes them into TOD. The design is flexible and can be customized to different instruments 
and survey designs. In the following, we describe the setup of the HIDE pipeline. This is not an exhaustive list for 
generic surveys and can easily be extended. The left half of Fig. \ref{fig:flow} shows the structure of the pipeline and 
includes the following steps:

\begin{description}

\item[Initialize] The \ivy configuration is loaded into memory and the random seed is initialized to ensure 
reproducible results.

\item[Load beam profile] A beam response pattern is loaded according to the configuration. The 
current design supports parametrized Gaussian or Airy \citep{Airy1838} profiles and arbitrary beam 
patterns specified on a grid. The beam profiles can be frequency-dependent.

\item[Load survey strategy] The pointing of the telescope at a given time in the survey is computed according 
to the desired survey strategy. The plugin converts the telescope pointing from terrestrial coordinates (azimuth, 
elevation) into equatorial coordinates  (RA, Dec). Different strategies can be chosen such as drift-scan 
surveys\footnote{A drift-scan is the survey mode where the telescope is fixed in the period of one day and changes 
the pointing of the telescope (mainly its elevation) from day to day. Due to the earth's rotation, the telescope 
footprint on the sky ``drifts'' through the full 2$\pi$ of RA positions at a given Dec on the celestial sphere.}, or a 
file-based scanning schedule  such that a planned survey can be exactly simulated.

\item[Load astronomical signal] The astronomical signal used for the simulation is loaded. Here we use 
the Global Sky Model \citep[GSM,][]{deOliveira-Costa2008}, which is a synthetic model of the Milky Way 
as a function of frequency based on a large number of radio data sets. 
The GSM is stored as \healpix maps on a grid of frequencies and interpolated to the 
desired frequency when needed. The GSM maps are in units of brightness temperature (Kelvin). 

\item[Convolve signals with beam] For each telescope pointing defined by the survey strategy, the 
telescope beam response is convolved with the astronomical signals and appended to the TOD array.

\item[Apply gain] To transform the TOD from units of Kelvin into internal units (Analog-to-digital unit, 
ADU) as recorded in the instrument, we multiply the TOD by a gain template. This information can 
come from external calibration or specifications of the instrument. 

\item[Add baseline] An frequency and point dependent baseline offset is added to
the TOD to account for contributions to the overall intensity from the instrument, the environment, etc. (see Section \ref{sec:simulatingbgs_hide} for more details)

\item[Add RFI] A simulated RFI signal is added to the TOD array (see Section \ref{sec:simulatingbgs_hide} for 
the specific model derived from data).

\item[Add noise] Gaussian (white) noise and and $1/f$ (pink) noise is added to the data to model noise 
contribution from the atmosphere and electronics.

\item[Write TOD] The simulations are written to disk in TOD format.
\end{description}

We note that the most computationally demanding step in the \hide pipeline is the step of convolving the 
signals with the beam. We speed up this step by using \textit{Quaternions} and a KD-tree data structure. 
The technical details of this operation is described in \ref{sec:beam_convolution}. 

\subsection{\seek architecture}
\label{sec:seekarchitecture}

\seek is a flexible and easy-to-extend data processing pipeline for single dish radio telescopes. It takes 
the observed (or simulated) TOD in the time-frequency domain as an input and processes it into \healpix 
maps while applying calibration and automatically masking RFI. The data processing 
is parallelized using \ivy's parallelization scheme.

We outline the setup of the \seek pipeline below. Again, this list is not exhaustive but can easily be extended 
and modified given a different experiment. The structure of the \seek pipeline is illustrated in the right half of 
Fig. \ref{fig:flow} and includes the following steps:

\begin{description}

\item[Find files] The file system is traversed to find simulated or observed data for a given time period and 
file-name convention. 

\item[Load data] The data is loaded from the file system into memory and smoothed 
if specified by the user. \seek is currently able to process both FITS \citep{Wells1981} and 
HDF5\footnote{\url{https://www.hdfgroup.org/HDF5}} data formats. The design concept of \ivy ensures 
that the other plugins do not depend on the origin of the data. Therefore
extending the support for further file formats can be implemented without
interfering with the other functionalities.

\item[Apply gain] A gain factor is computed by using special calibration data that was collected on 
dedicated calibration days in the survey. Alternatively, an externally provided template can be loaded from 
the file system. This gain factor is applied to the TOD to convert the instrument-recorded values (ADU) to 
physical units (Kelvins).

\item[Coordinate transformation] The telescope pointing at each given time is connected to the TOD to 
give each pixel in the TOD a terrestrial coordinate. These coordinates are now transformed into equatorial 
coordinates corresponding to a given point in the celestial sphere.

\item[Mask objects and bad frequency channels] The TOD is masked if known
bright objects such as the Sun and the Moon are too close to the telescope
pointing. Furthermore, frequency bands known to be unusable (e.g., seriously
contaminated by satellite communication bands) are masked.

\item[Mask RFI] The TOD is analyzed and pixels identified as contaminated by RFI
are masked (see \ref{sec:rfi} for more details on \seek's automated RFI masking
mechanism).

\item[Baseline removal] The baseline level per frequency is estimated from the median value over time of 
the cleaned TOD. This baseline is subtracted from the TOD. 

\item[Create map] For every frequency channel the TOD is processed into a \healpix map. By default, each 
pixel in the \healpix map is filled with the mean value of all measurements in that pixel. One can also invoke 
an outlier-rejection step to avoid uncleaned RFI contaminating the mean. 

\item[Write map] The \healpix maps and auxiliary information such as the frequency range and 
redshift information is written to disk in HDF5 file format.

\end{description}

In the \seek pipeline, tackling the RFI is the most computationally challenging task. This includes 
both the RFI masking in the TOD plane and the final outlier-rejection applied at the \healpix map 
level. We describe our specific treatment of RFI in \ref{sec:rfi}.

\subsection{Quality assurance}
\label{sec:quality}

We developed \hide \& \seek using best practice from software engineering. In particular, we
use tools common in the Python community. Both packages follow a standardized packaging, which 
simplifies the development and installation including resolving dependent third party packages. The 
standardization released by the Python Package authority defines the directory
structures of a package, which enforces the separation of functionality and makes it easier for new developers to 
engage in the project. Furthermore, it defines how to store meta-information of the package.
In order to maintain a high level of quality and that newly developed features do not infer with existing 
code, we rigorously test the functionality of the packages with unit tests. To develop those tests we use 
the common testing framework {\texttt{py.test}\xspace}. Finally, both packages are fully documented 
using the standardized {\texttt{reStructuredText}\xspace} syntax such that we can automatically 
generate and publish a documentation\footnote{\url{https://hide.readthedocs.io}}\footnote{\url{https://seek.readthedocs.io}} using the {\texttt{Sphinx}\xspace} package.

\section{Forward-modeling a Galactic survey at 990 -- 1260 MHz with \hide \& \seek}
\label{sec:application2bgs}

In this section, we describe how we apply \hide \& \seek to model and analyze data for a mock 
Galactic survey. The design of this mock survey is based on the early Science Verification (SV) data from 
the Bleien Observatory as described in C16 and allows us to develop realistic models and test the 
robustness of the code. This data set was collected using a 7m single-dish telescope operating in 
drift-scan mode with a frequency range of 990 -- 1260 MHz. In addition to the continuous scanning, 
calibration data was taken every 8 days to provide an anchor for various calibration tests. For more 
details of this data set, we refer the readers to C16.

We discuss below the survey-specific implementations that we have adopted in \hide
(Section~\ref{sec:simulatingbgs_hide}) and \seek (Section~\ref{sec:simulatingbgs_seek}), respectively. For 
other experiments, these implementations may need to be modified or replaced. We then show an example of a 
forecast for the Galactic survey in Section~\ref{sec:forecast} with the current pipelines.

\subsection{Simulating a Galactic survey with \hide}
\label{sec:simulatingbgs_hide}

We customize the functionalities of \hide to match the characteristics of the SV data from the Bleien 
Observatory. For the beam profile, or the angular response of our telescope, we use an Airy disk profile, 
which is a good approximation for an underilluminated 7m parabolic dish (C16). 
For the size of the beam, we use values inferred from transits of the Sun as described in C16. 
We generate data with $\sim1$ MHz bandwidth between 990 MHz and 1260 MHz, resulting in 276 
frequency channels. We use a drift-scan strategy that follows the schedule used in the SV data. 

\begin{figure}[t]
\begin{center}
\includegraphics[width=0.4\linewidth]{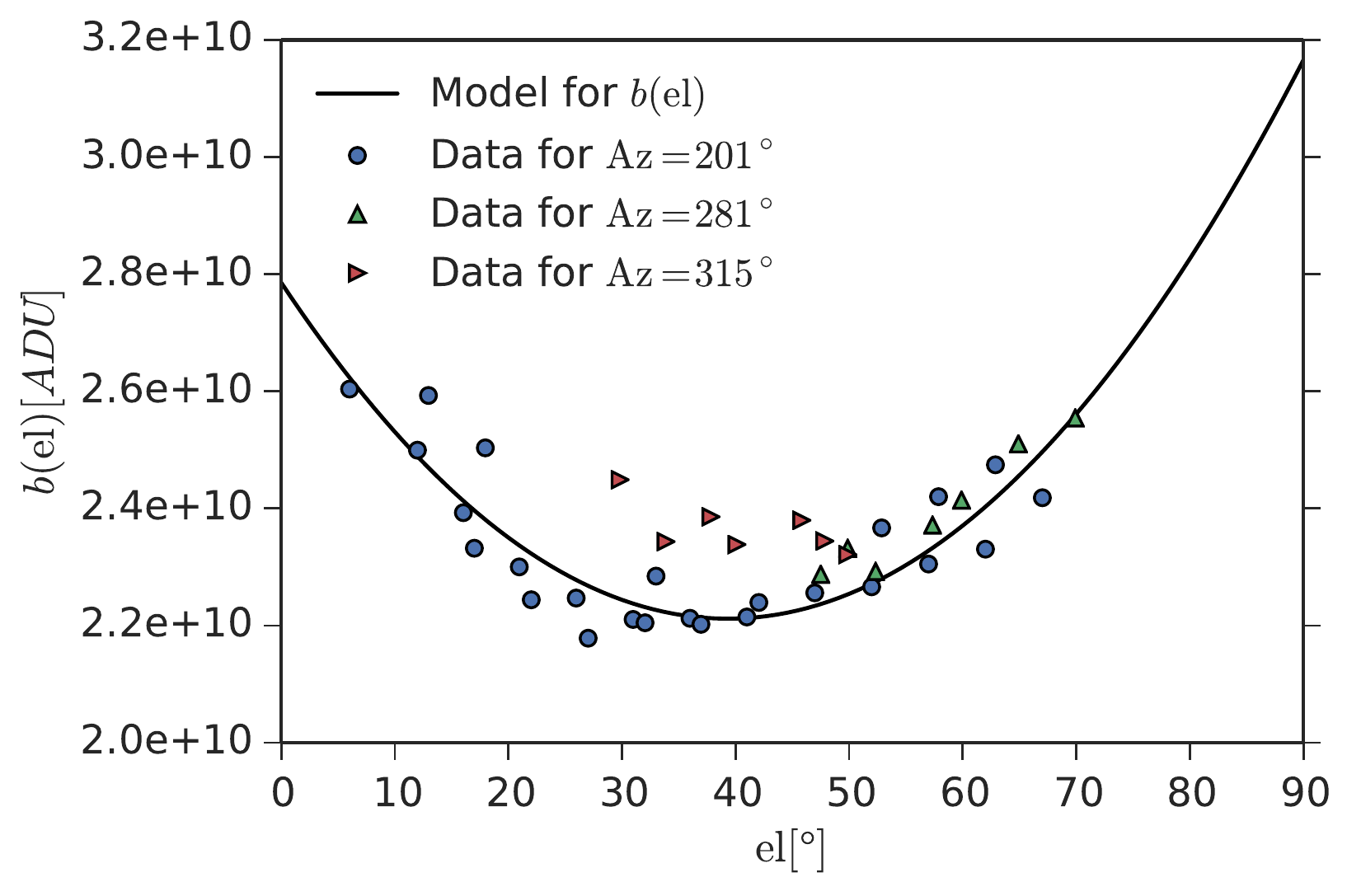}
\includegraphics[width=0.43\linewidth]{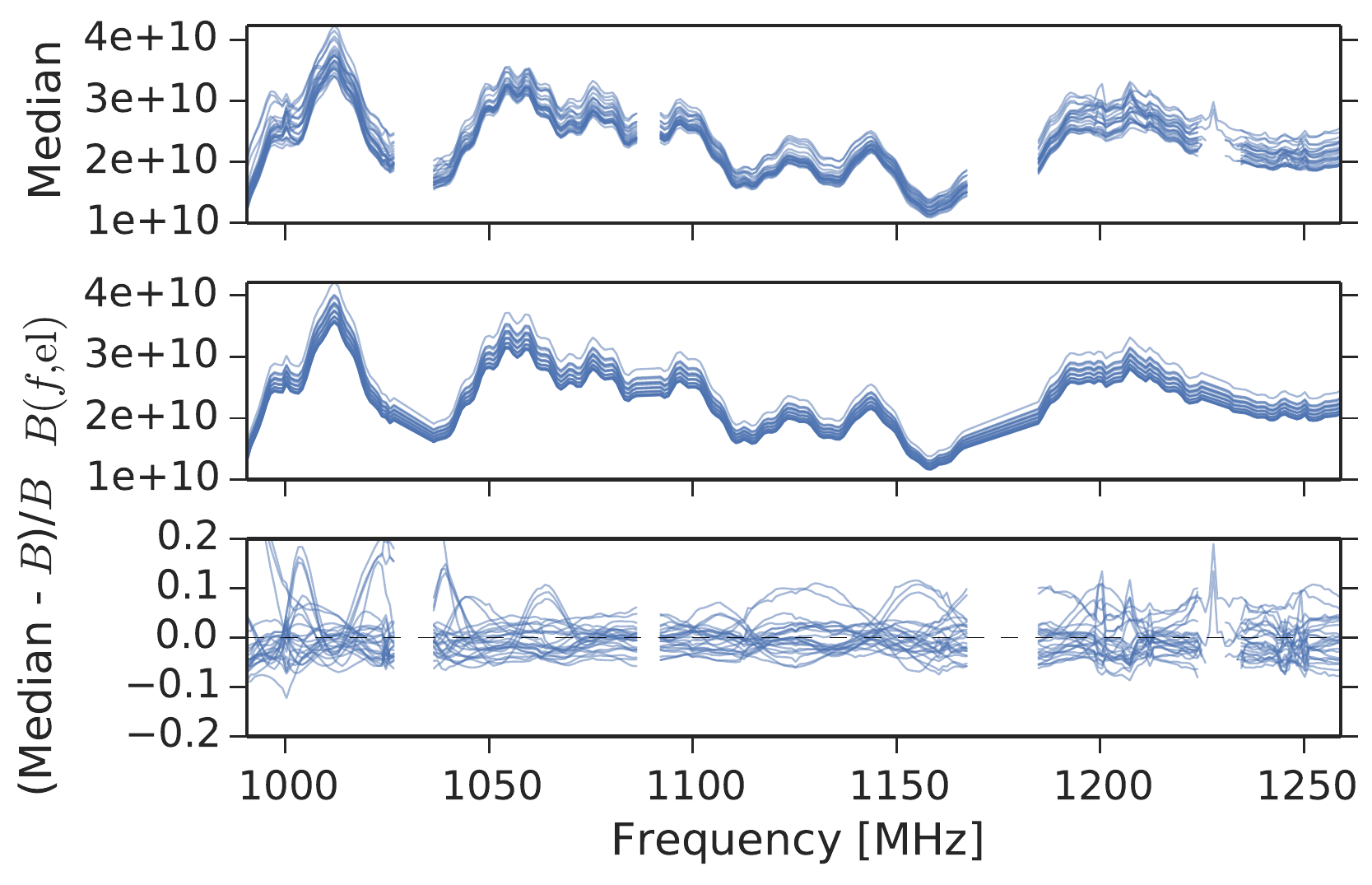}
\caption{The left panel shows the baseline model (black line) as a function of elevation. The model is 
a second order polynomial fitted to SV observation at different elevations and azimuth (green, blue and 
red dots). The right panel depicts the comparison of the median of the observed TOD in SV and 
the baseline model $B(f, el)$ used for the forecast.}
\label{fig:elmodel}
\end{center}
\end{figure}

We make the simplified assumption that the TOD has a frequency- and
elevation-dependent baseline that is otherwise independent of azimuth and time of the pointing as well as its Galactic coordinates. We furthermore 
assume that the frequency- and elevation-dependencies factorize, i.e. our baseline $B(f, el)$ 
is given by
\begin{equation}
	B(f, el) = b(f) \delta b(el),
\label{eq:baseline}
\end{equation}
where $f$ is the frequency, $el$ is the elevation, $b(f)$ is the frequency-dependent baseline and 
$\delta b(el)$ is a further elevation-dependent modulation of the baseline.
To generate the model parameters for Eq.~\eqref{eq:baseline} that are motivated by data, we calculate the 
median of the TOD over a time window during the night (from midnight to 4:00 UT) over all survey days 
during SV after RFI mitigation. We then fit a second-order polynomial to the mean of the medians as a 
function of elevation (see Figure \ref{fig:elmodel}, left panel). Finally, we use the mean of the normalized medians 
as a model for the frequency-dependent baseline $b(f)$. In the right panel of Figure \ref{fig:elmodel}, we show the 
agreement between the observed medians and the calibrated model $B(f, el)$. We
note that this is a simplified model. In \ref{sec:comparison} we discuss the
influcence of day time and telescope pointing on the baseline for the SV data
from the Bleien Observatory.

As shown in C16, the instrument noise in our TOD is very close to white noise due to 
the stable phase-switch implemented in the system. To model the amplitude of the noise, we estimate the 
standard deviation from TOD recorded in the night during SV. 
To avoid biases from RFI that was not properly mitigated, we estimate the scale of the noise 
per frequency channel as the mean of the $16$ and $84$ percentiles of the TOD for a given night. We 
use the mean of the estimated scales over all days for clean channels and the minimum over all days for 
the contaminated channels as model for the standard deviation of the TOD in units of ADU.

\begin{figure}[t]
\begin{center}

	\minipage{0.5\textwidth}
	  \includegraphics[width=\linewidth]{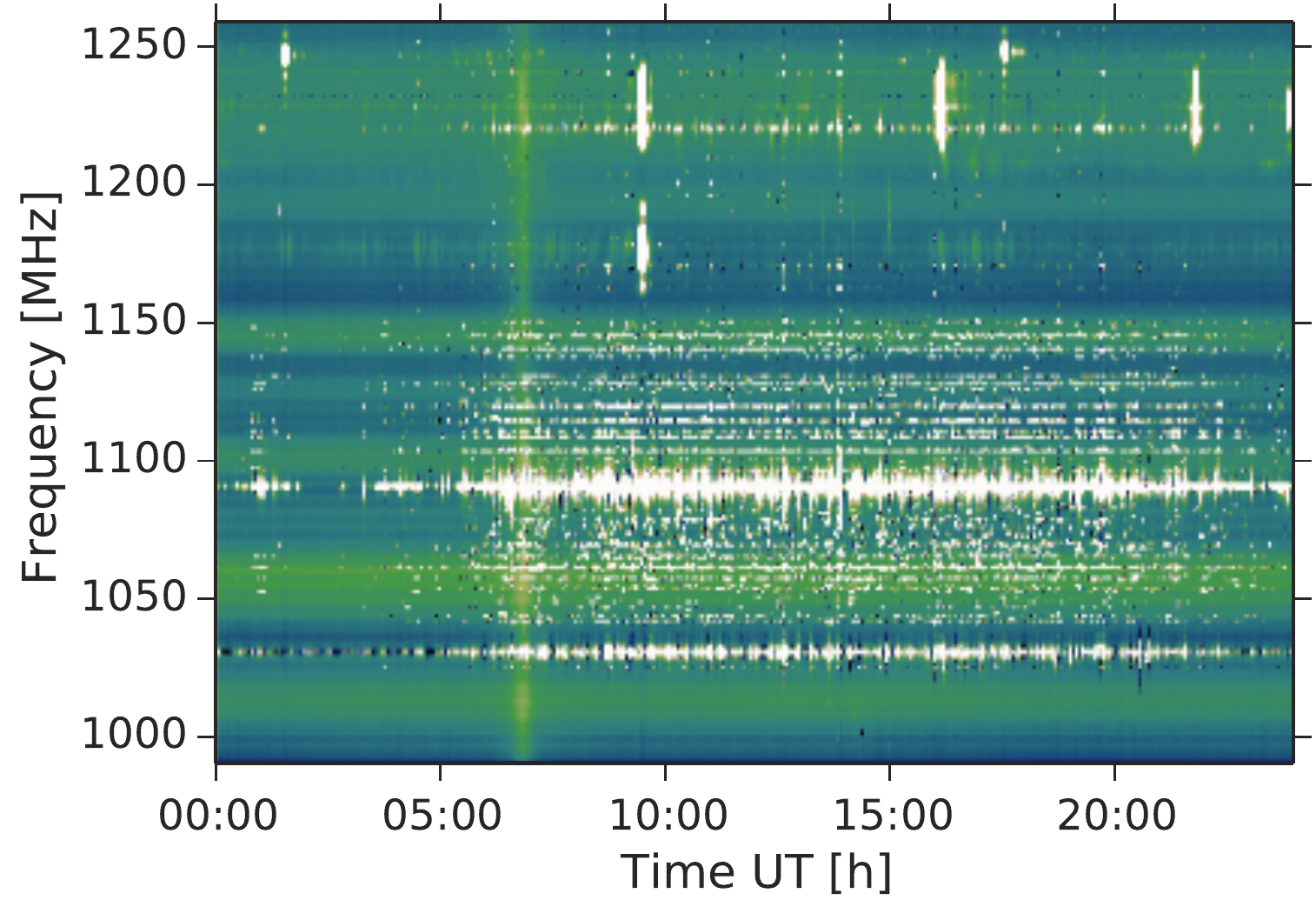}
	\endminipage\hfill
	\minipage{0.5\textwidth}
	  \includegraphics[width=\linewidth]{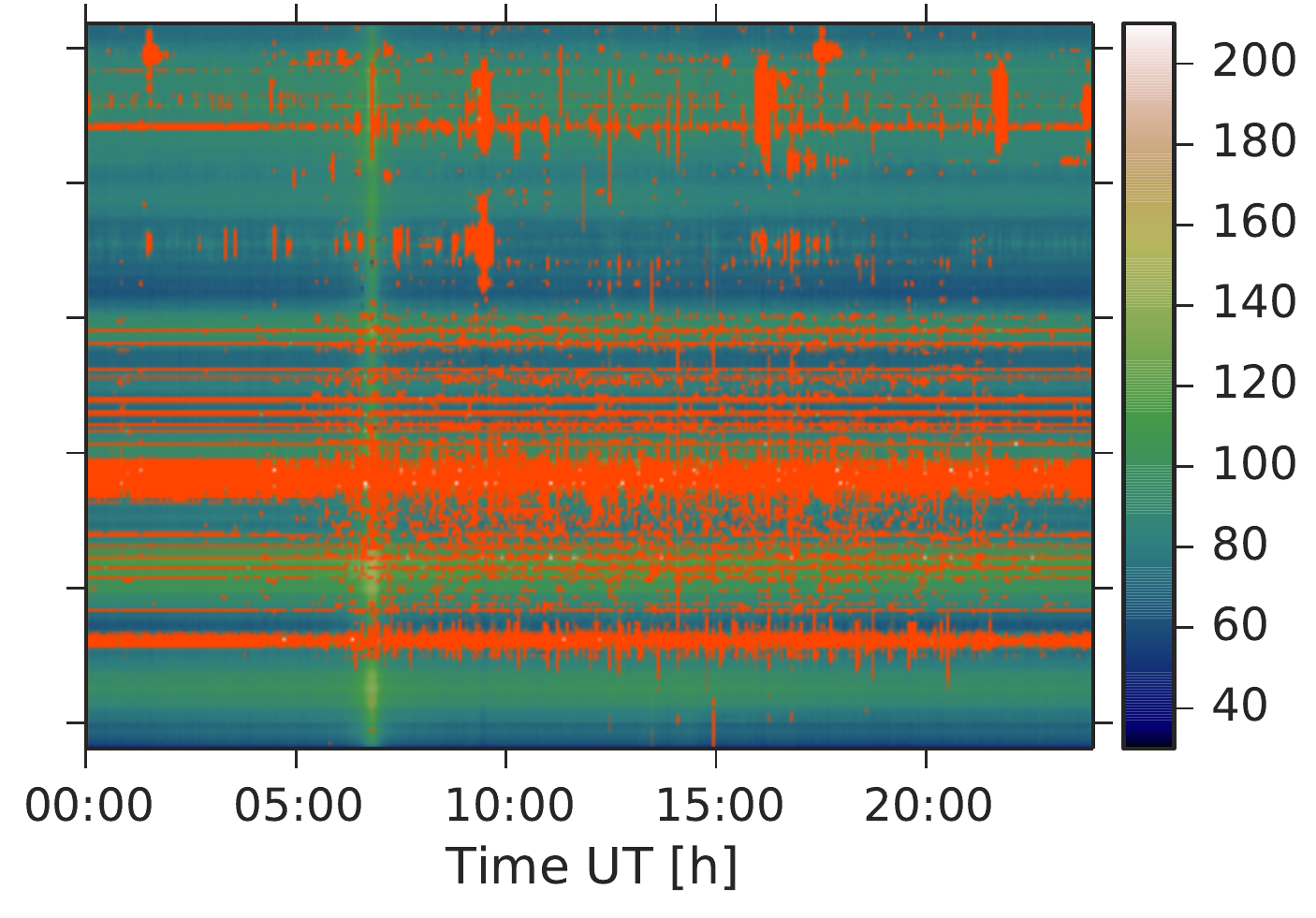}
	\endminipage\hfill
	\minipage{0.5\textwidth}
	  \includegraphics[width=\linewidth]{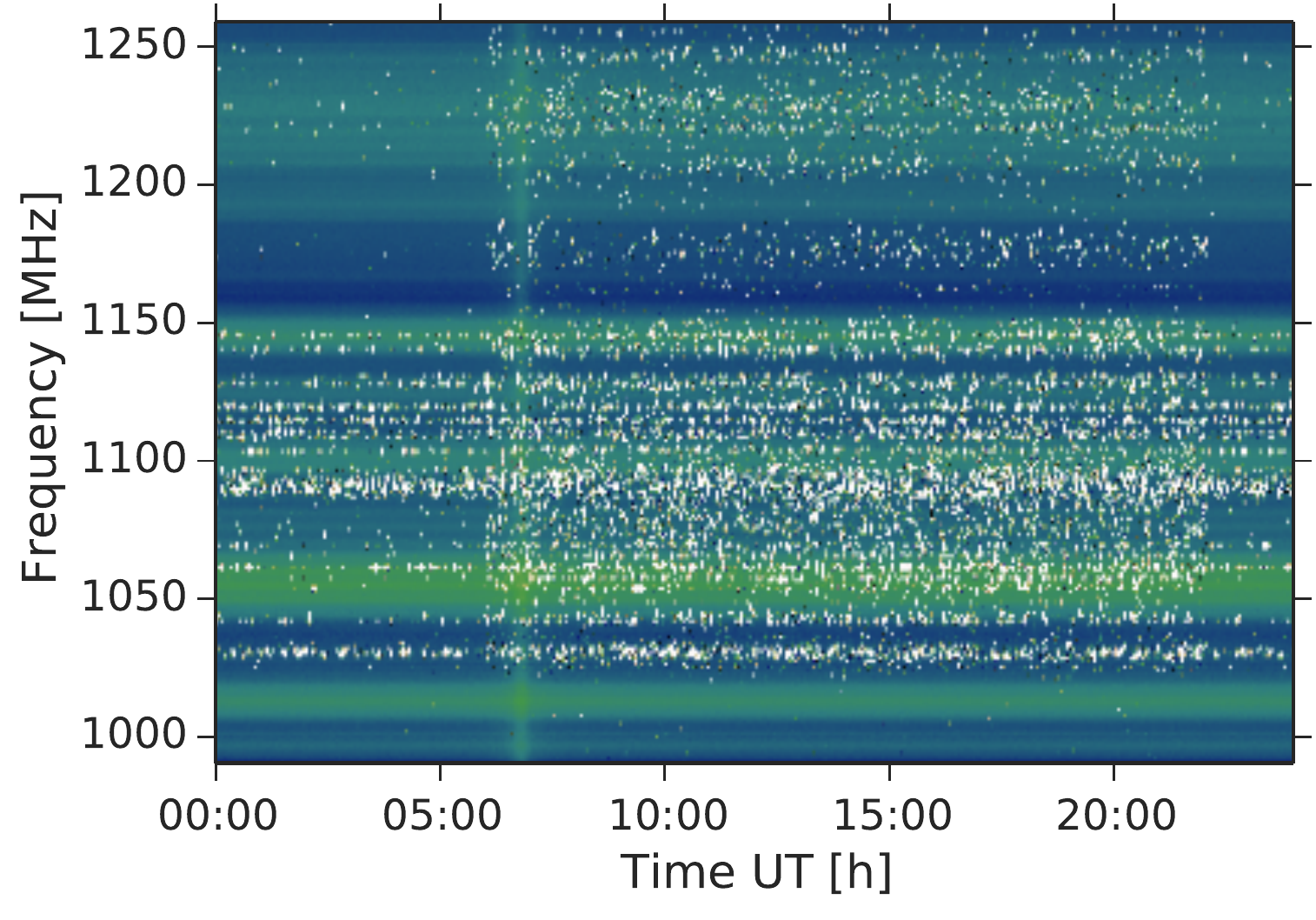}
	\endminipage\hfill
	\minipage{0.5\textwidth}
	  \includegraphics[width=\linewidth]{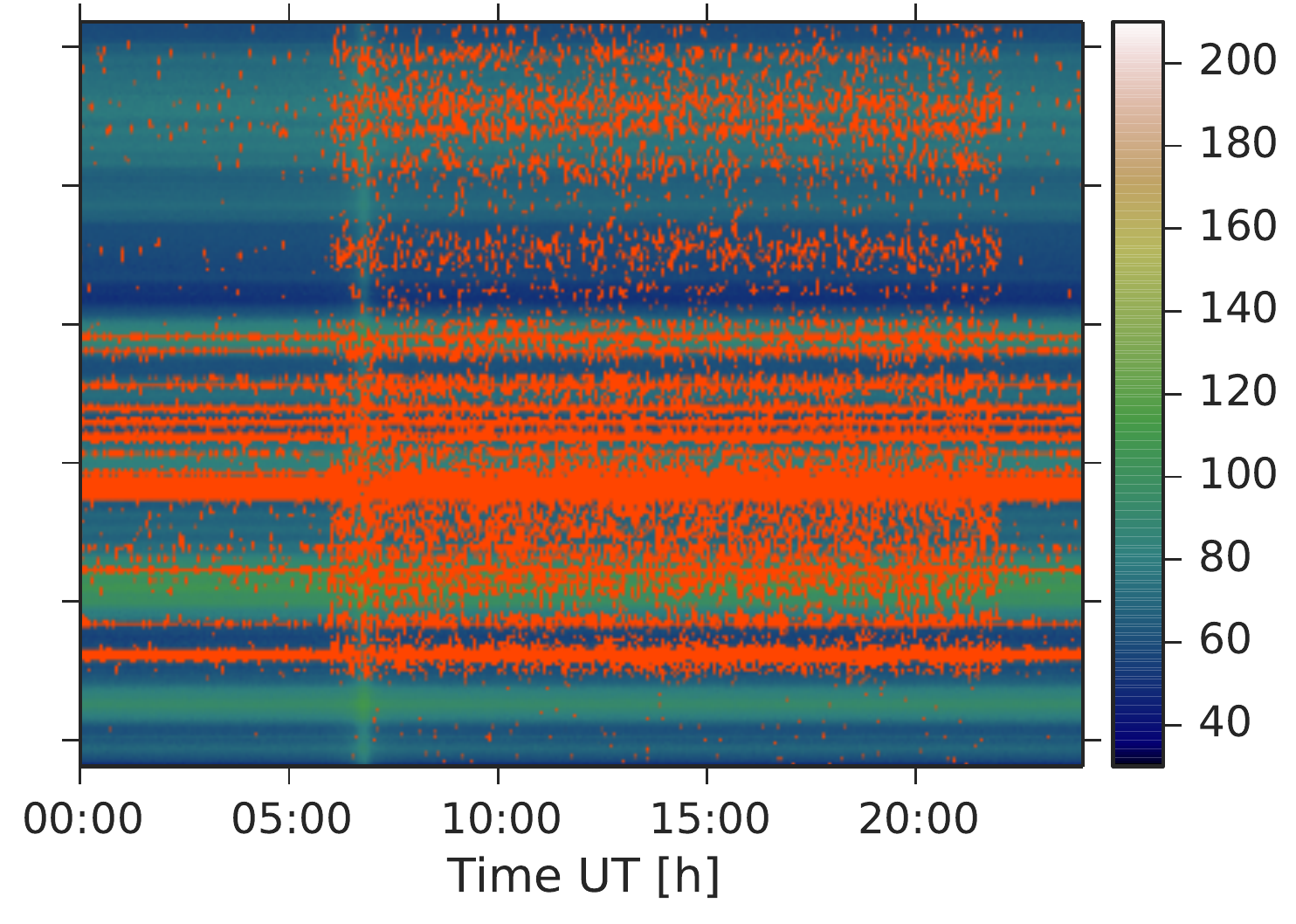}
	\endminipage\hfill

\caption{The left panel in the top row displays the unprocessed TOD recorded on
21st of March 2016. The broadband RFI contamination mainly coming from the nearby airport and is visible in the 1025--1150 MHz frequency 
band. The TOD also shows the variation between day and night as the amount of RFI increase at around 6:00 am 
and decreased at 11:00 pm UT. The the lower left panel shows the TOD simulated
with \hide. The right pannels show the corresponding TOD with the RFI mask overlaid (orange).}
\label{fig:tod}
\end{center}
\end{figure}

The Bleien Observatory is in a location with non-negligible RFI from human
produced sources such as satellites and aircraft communication in the frequency
range of interest. Due to the large variety of distinct RFI patterns, developing a realistic RFI model is challenging. 
As a result, we choose to adopt a simple model for the RFI, where each RFI burst is defined 
by the same profile in time and frequency (either exponential or Gaussian) and an amplitude. For each frequency, 
the amplitudes are sampled randomly from a uniform distribution between the standard deviation of the noise 
and a maximum value inferred from the real data. The rate of RFI bursts per frequency is chosen such that, on 
average, the number of pixels affected by RFI greater than the noise-level is matching the fraction of masked pixels 
inferred from the real data. This model can be implemented efficiently by starting with a zero time-frequency 
plane and setting the randomly chosen positions in time to the randomly drawn amplitudes for each frequency. 
The final simulated RFI is then given by a two-dimensional FFT convolution of the time-frequency plane with the 
chosen RFI profile. Figure \ref{fig:tod} shows the TOD from the Bleien
Observatory (first row) on March 21st, 2016, next to a simulated TOD from \hide
(second row).
We note that even though this model is not capturing all aspects of the RFI, it approximately reproduces the 
RFI-induced data-loss and gives us the possibility to explore the performance of the RFI mitigation in \seek. 

\subsection{Analyzing simulated data with \seek}
\label{sec:simulatingbgs_seek}

Next, we analyze the TOD and reconstruct the astronomical signals with \seek. 
Having a simulation that is expected to be close to the observation allows us to implement a \seek pipeline 
that can be used for the processing of the observation as well as simulation data with minimal modifications.  

\begin{figure}
\begin{center}
\includegraphics[width=0.6\linewidth]{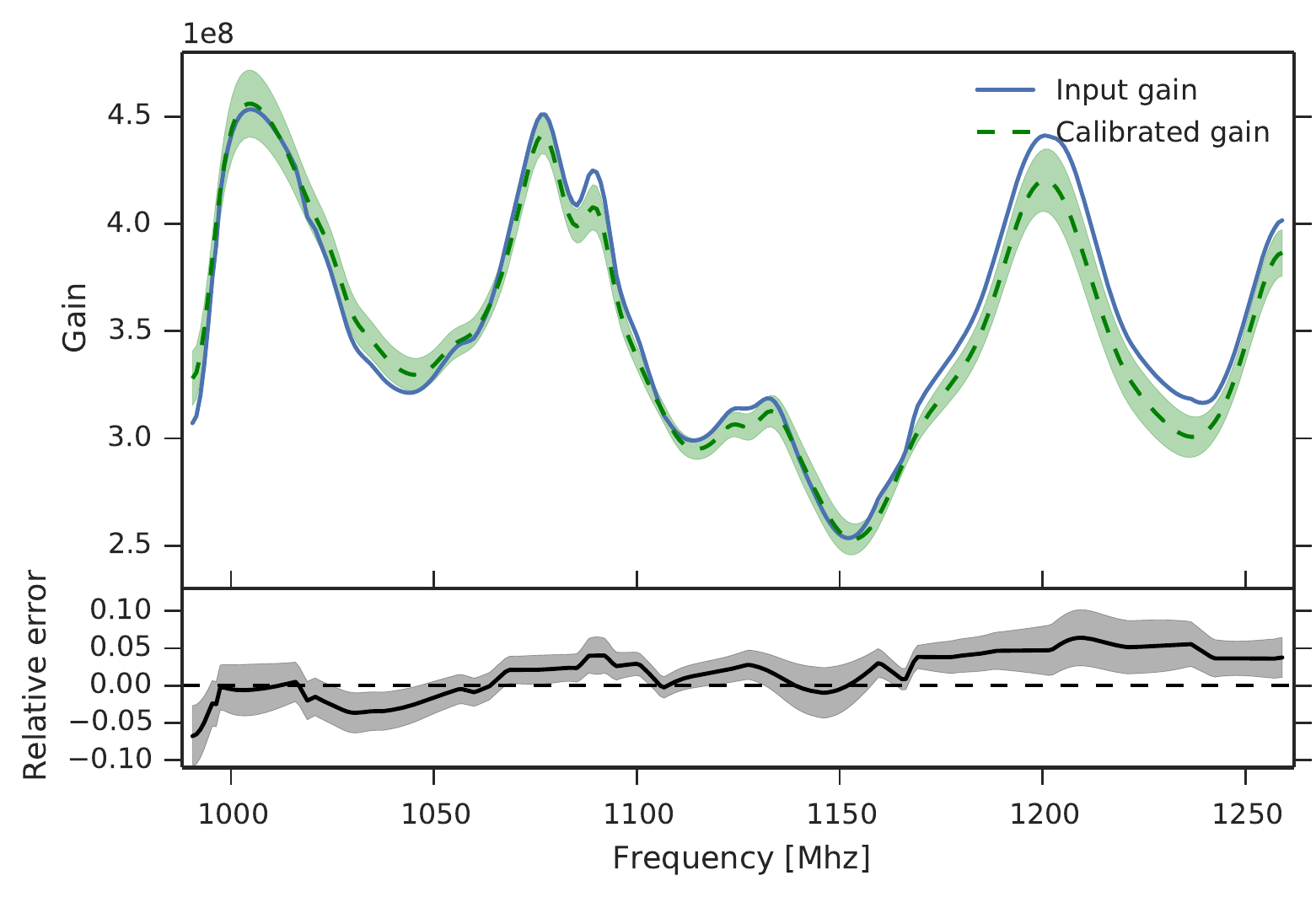}
\caption{Input gain from the \hide simulation (blue) and the derived gain calibrated form 16 calibration days 
in \seek (green) as a function of the frequency. The green solid line is the median and the range shows the 
one standard deviation spread over all calibration days. The lower panel shows the relative error of the 
calibration process.}
\label{fig:gain}
\end{center}
\end{figure}

The first step after loading the TOD in the \seek pipeline is to obtain the gain factor $G(\lambda)$ that 
converts the observed ADU values in the TOD into physical units. This procedure is referred to as flux 
calibration and the exact implementation is described in \ref{sec:flux_calibration}. Fig. \ref{fig:gain} shows 
the gain factors used for the simulation in \hide (blue line) and the gain factor 
recovered from \seek's flux calibration. The accuracy of the flux calibration is biased by the RFI contamination 
and the noise in the data set. We use a simplified model and assume that $G(\lambda)$ is constant over 
time. From repeated measurements in SV we see variation in $G(\lambda)$ can vary up to $10\%$. 

\begin{figure}
\begin{center}
\includegraphics[width=0.6\linewidth]{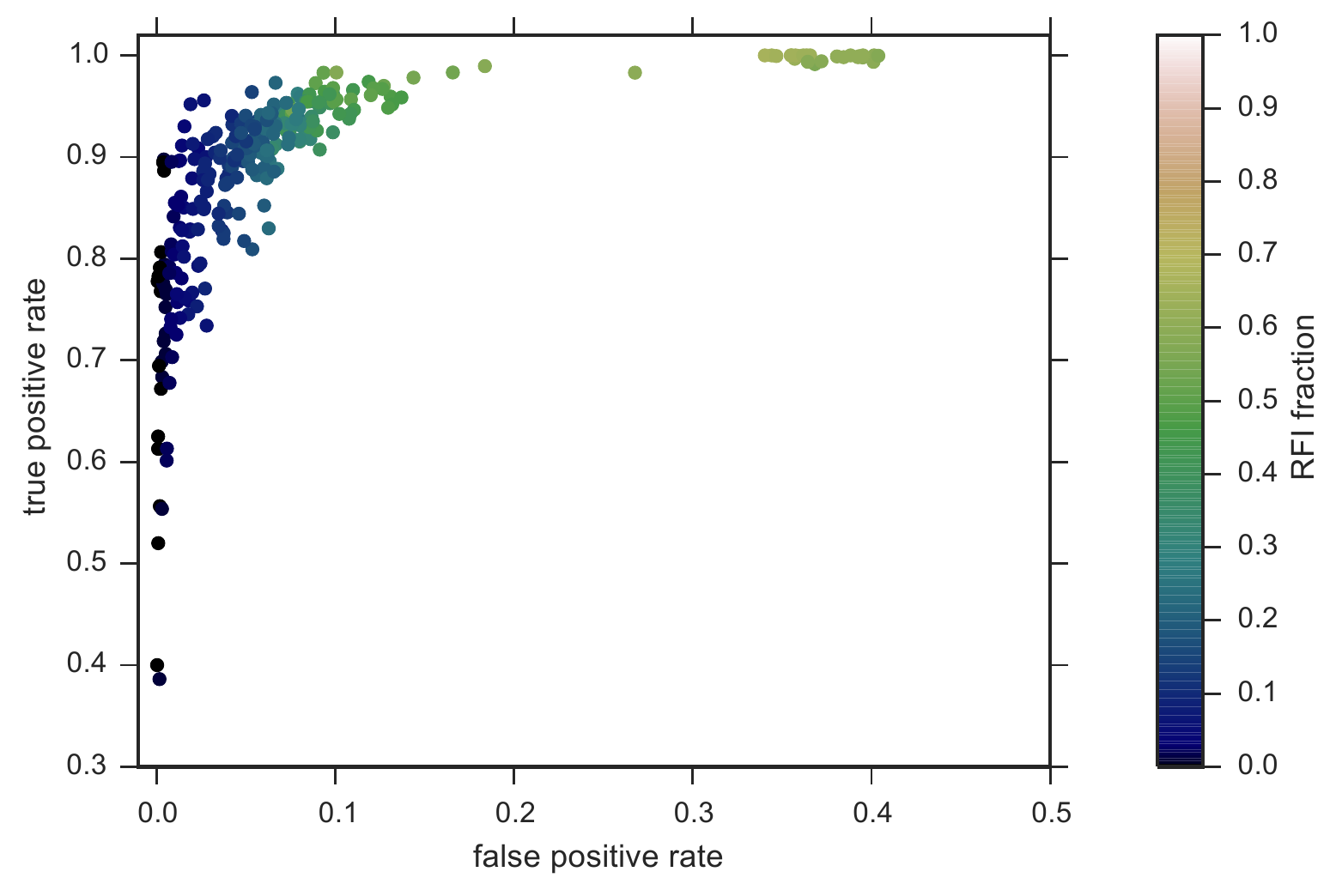}
\end{center}
\caption{The receiver operating characteristic (ROC) curve shows the ratio of pixels correctly identified as RFI 
pixels (true positives) as a function of the ratio of pixels incorrectly masked (false positives). Each point 
represents one frequency channel (with a bandwidth of $\sim1$MHz). The color scheme displays the amount 
of RFI present in one channel. If a channel suffers from little RFI, \textsc{SumThreshold} misses some of the contaminated 
pixels. As soon as the RFI fraction increases most of the pixels are correctly identified as unwanted signal.}
\label{fig:roc}
\end{figure}

For object-masking, we configure \seek such that data is masked if the angular separation of 
the telescope pointing and the Sun is smaller than $15^{\circ}$. Depending on the Moon phase data is 
masked if the separation is smaller than $5-10^{\circ}$. 

For RFI mitigation we have implemented the \textsc{SumThreshold} algorithm \citep{offringa2010post} 
with a morphological dilation of the resulting mask (for more details we refer to \ref{sec:rfi}). We set the 
parameters for the \textsc{SumThreshold} and mask dilation such that most of the RFI is being masked while 
minimizing the fraction of incorrectly masked data. The tuning of these parameters is done by analyzing 
the RFI masks of simulated data where the positions of the RFI pixels are exactly known. We then 
inspect the mask visually and statistically. 
Fig. \ref{fig:roc} shows the receiver operating characteristic (ROC) curve of the mask obtained from 
\textsc{SumThreshold} and mask dilation. The x-axis displays the ratio of pixels incorrectly labeled as RFI 
and the y-axis represents the ratio of correctly masked pixels. Each point represents a different frequency while 
the color is showing the fraction of RFI in this channel. If the RFI contamination is low in a specific channel, 
\textsc{SumThreshold} has a low probability to mask the RFI pixels. As the fraction of contaminated pixels 
increases, the algorithm correctly masks the unwanted signal while maintaining a low rate of incorrectly masked 
clean pixels. In highly RFI-contaminated channels the true positive rate reaches almost $1$. However, due to the 
high density of masked pixels, the mask dilation also increases the false positive rate.

Fig. \ref{fig:tod} displays the visual inspection of a representative observed
and simulated data set recorded on 21st of March 2016. The two left panels (observed and simulated TOD respectively)
show the unmasked data where strong broad- and narrow-band RFI contamination is
visible. The right panels show the same data overlaid with the RFI mask obtained
from \textsc{SumThreshold} and mask dilation. Most of the RFI is well captured
including weaker narrow-band RFI not visible in the left panels.

\begin{table}[t!]
\caption{Description of \hide \& \seek parameters used in forecasting of the Galactic survey.}
\begin{center}
\begin{tabular}{lll}
\hline
\hline
&Name & Value \\
\hline
\multirow{10}{*}{\hide} & {\tt Scanning strategy} & Drift scan\\
&{\tt Start date} & 16.12.2015\\
&{\tt End date} & 26.05.2016\\
&{\tt Telescope latitude} & 47.344$^{\circ}$\\
&{\tt Telescope longitude} & 8.114$^{\circ}$\\
&{\tt Dish diameter} & 7 m\\
&{\tt Beam profile} & Frequency-dependent Airy disk\\
&{\tt Astronomical signal} & Global Sky Model with artificial calibration sources\\
&{\tt Near earth signal} & Bleien horizon model (Fig.~\ref{fig:elmodel}) \\
&{\tt RFI model} & Bleien RFI model (Fig.~\ref{fig:tod})\\
\hline
\hline 
\multirow{7}{*}{\seek} &
\multirow{2}{*}{\tt Sum threshold} & $\chi=10$, smoothing kernel:40$\times$20 pixels \\
& &$\sigma_i=15$, $\sigma_j=7.5$ \\
& {\tt Mask dilation} & Binary dilation with $6 \times 6$ pixel structure\\
&{\tt \healpix map making} & Pixel averaging with outlier rejection\\
&{\tt Minimum Sun separation} & $15^{\circ}$\\
&{\tt Minimum Moon separation} & $5-10^{\circ}$\\
&{\tt \healpix resolution} & Nside=64 (0.84 deg$^2$)\\
\hline
\end{tabular}

\end{center}
\label{tbl:params}
\end{table}

\subsection{Forecast}
\label{sec:forecast}

In this section we apply the full \hide \& \seek framework to produce an end-to-end simulation for the
mock Galactic survey described above. That is, given the survey design and our understanding of the 
data characteristics from the Bleien Observatory, we forecast the expected map of the Milky Way we 
will measure at the end of SV data collection after processing all the data through \seek. We also 
forecast the expected signal-to-noise ratio of the observed Milky Way at different positions on the sky.

\begin{figure}[!htb]
	\minipage{0.5\textwidth}
	  \includegraphics[width=\linewidth]{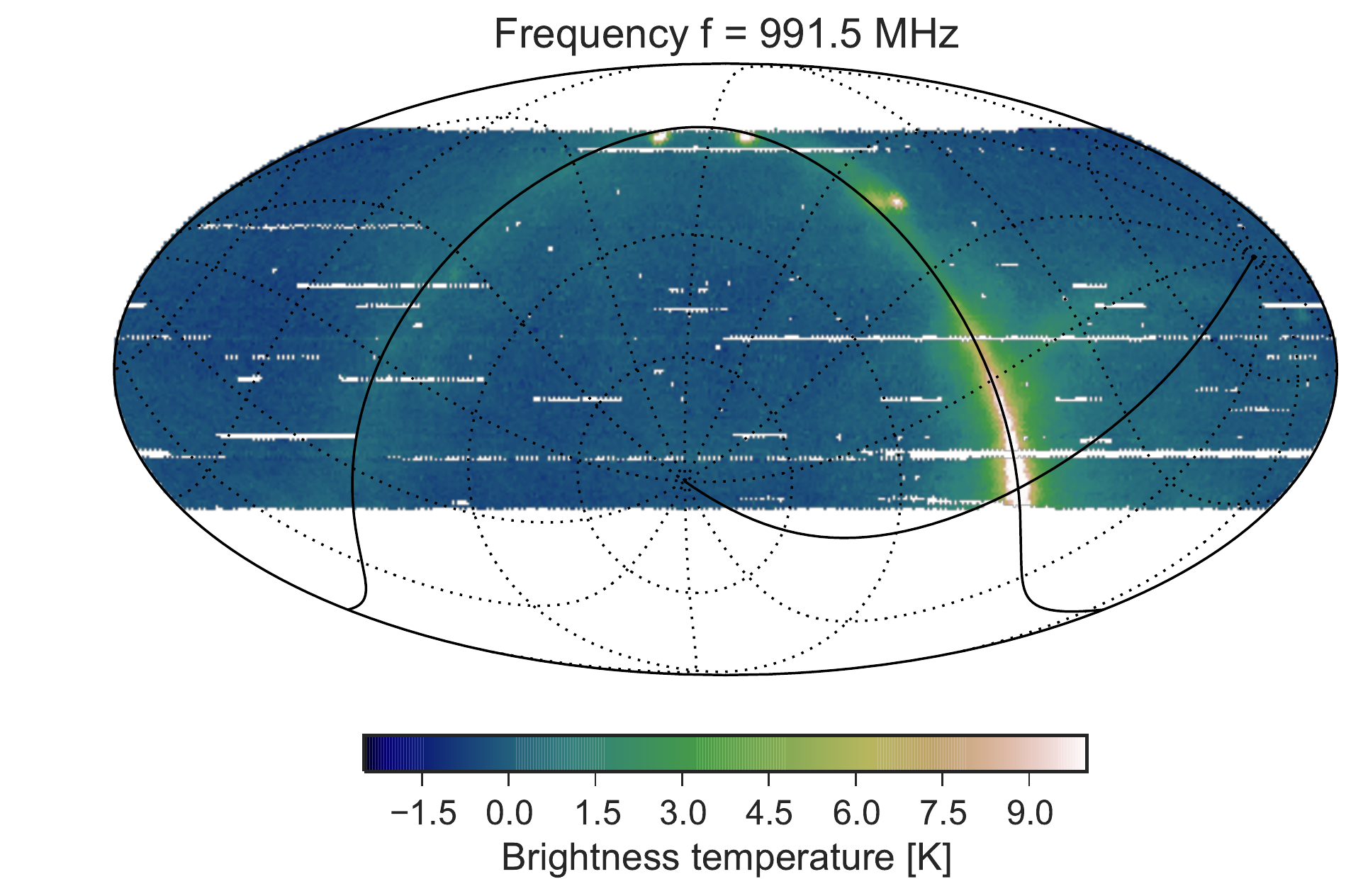}
	\endminipage\hfill
	\minipage{0.5\textwidth}
	  \includegraphics[width=\linewidth]{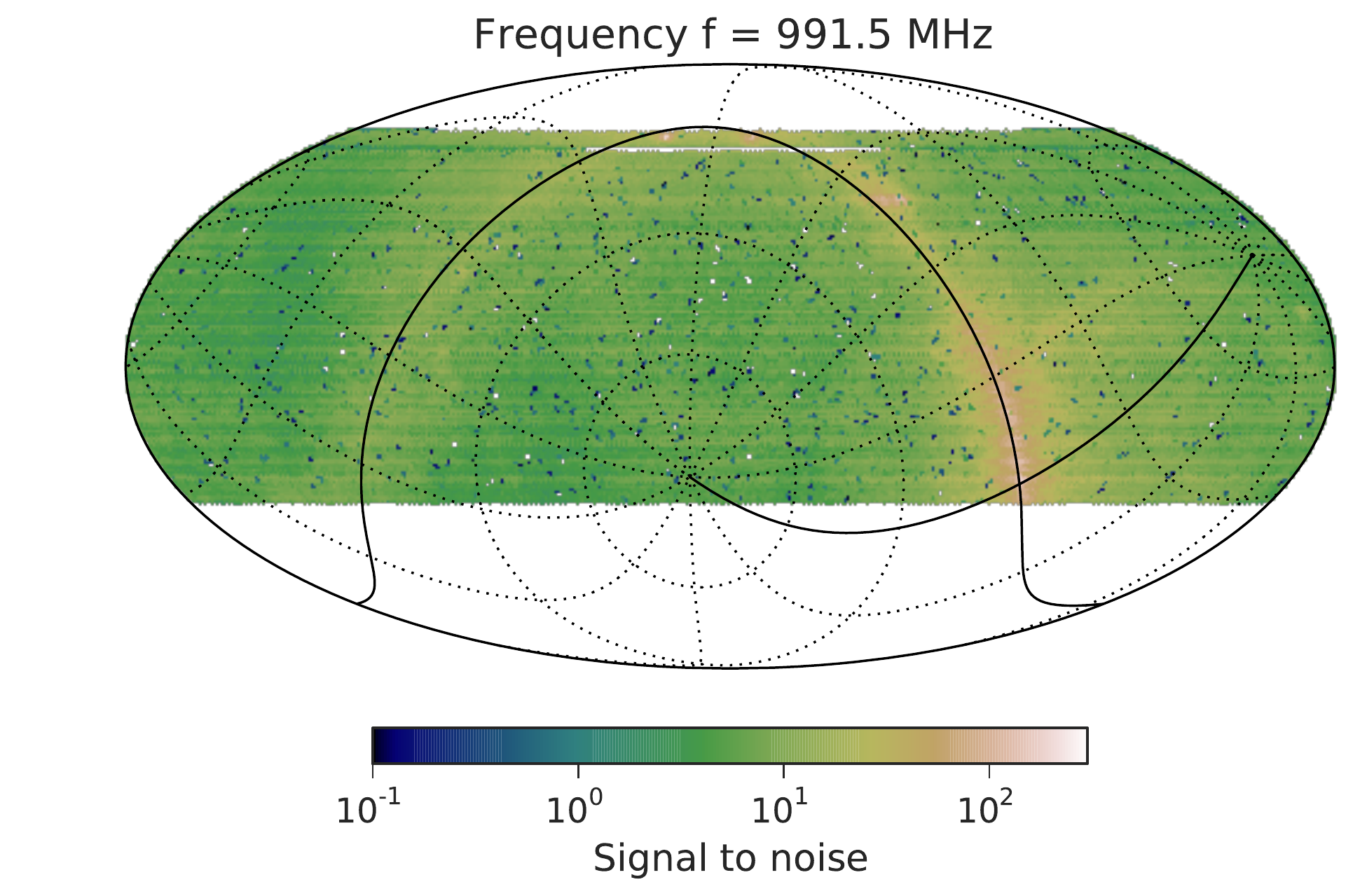}
	\endminipage\hfill
	\minipage{0.5\textwidth}
	  \includegraphics[width=\linewidth]{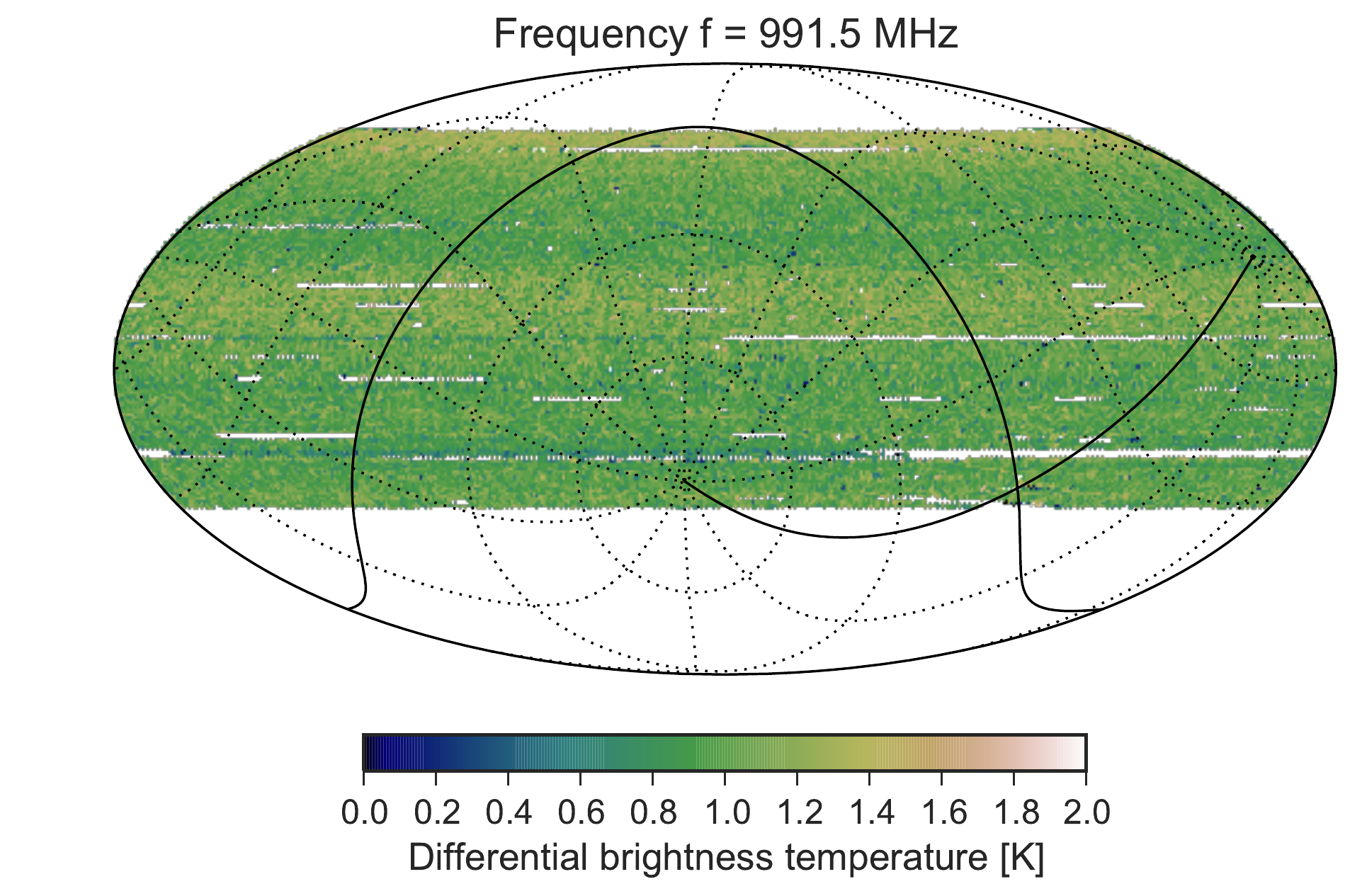}
	\endminipage\hfill
	\minipage{0.5\textwidth}
	  \hspace{0.3in}
	  \includegraphics[width=0.8\linewidth]{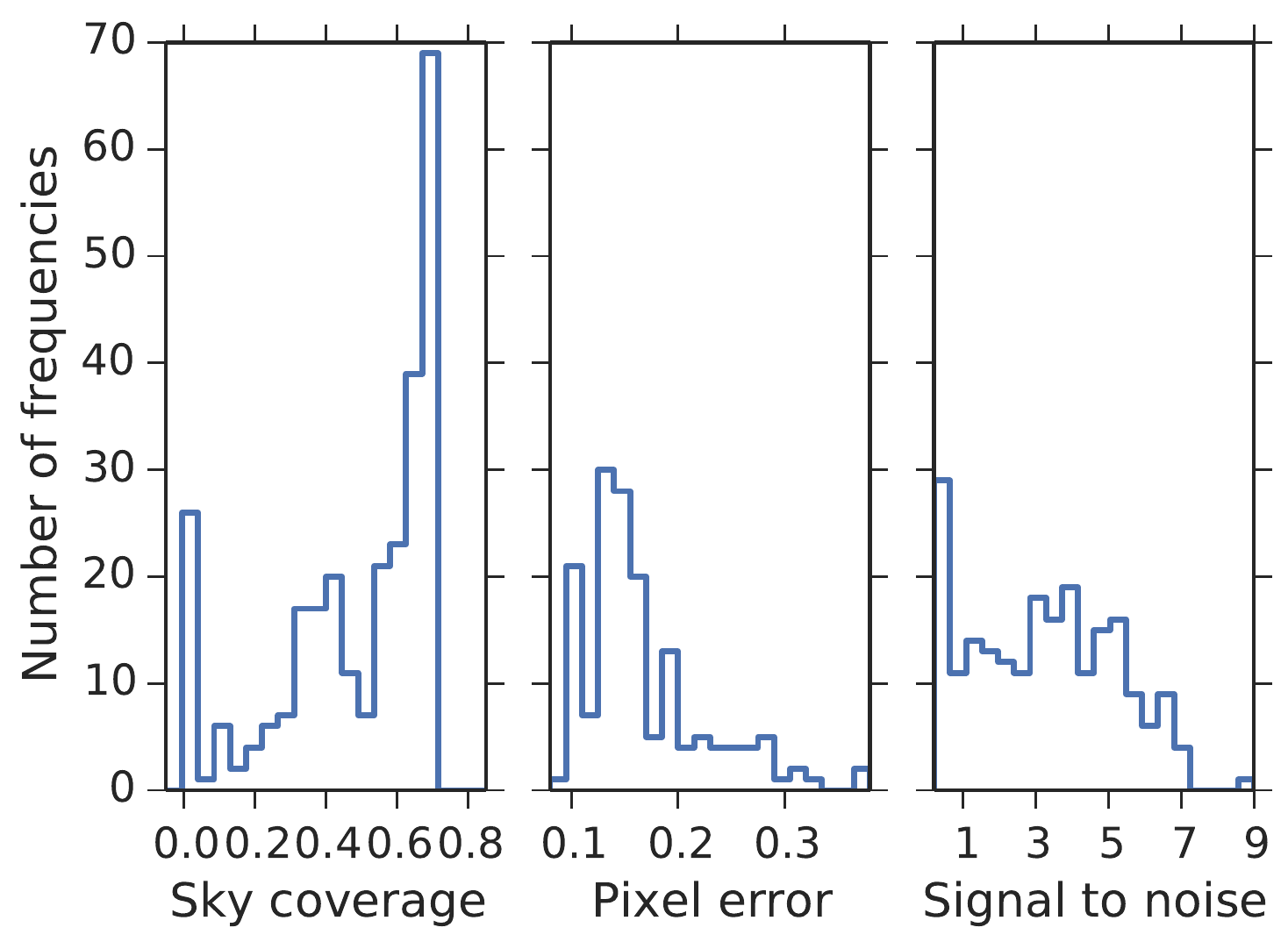}
	\endminipage\hfill
	\caption{Simulated maps at $f \approx 991.5$ MHz and statistics of the maps. Top left panel: 
	The forecast simulation including observational artefacts such as noise, RFI, gain and baseline. 
	Larger gaps in the map result mainly from the masking of the Sun and Moon. Despite being a 
	clean frequency channel, loss of the sky coverage due to RFI masking is visible as smaller masked 
	regions on the map. Top right panel: Map of the signal-to-noise ratio defined as the ratio between the 
	clean simulation and the forecasted noise. Lower left panel: Difference 
	between the clean map and the forecast map shown in the top left panel. Lower
	right panel: Distributions of the sky coverage, median noise, and median
	signal-to-noise of the final maps over all frequencies.}
	\label{fig:forecast}
\end{figure}

For this purpose, we set up a suite of \hide simulations with the parameters listed in the top half of 
Table~\ref{tbl:params}, and analyze the simulations using \seek with the parameters listed in the bottom 
half of Table~\ref{tbl:params}. 
These parameters are chosen to best match the SV data set from the Bleien Observatory.
We show examples of the resulting \healpix maps for this forecast run at 991.5 MHz 
in the top-left panel of Fig. \ref{fig:forecast}.
To obtain the pure ``signal'' for the signal-to-noise calculation described below, we also create a second 
\hide simulation without observational effects such as noise, RFI, gain, and baseline. Analyzing this second 
simulation with \seek yields a map that is convolved in the same way as the forecasted map, but does not 
suffer from any other observational effects.  

From the SV data, we inferred strong variations of the RFI contamination over frequencies as well as 
time of day. RFI from airplane communication, for example, only happens during the day as Zurich airport is 
open from 6:00 am to 11:00 pm UT and affects the TOD in the frequency range from
$1050$ MHz to $1150$ MHz (see Fig. \ref{fig:tod}, top left panel). As RFI leads
to data-loss, both the signal to noise and the sky coverage of our forecasts depend on frequency. We define 
signal-to-noise of a pixel in a given frequency channel as the ratio between the clean 
simulation (signal) and the uncertainty in the forecasted map estimated from the standard deviation in each pixel 
(noise). As an example, we show the signal-to-noise map at a frequency of 991.5 MHz in the top-right panel 
of Fig. \ref{fig:forecast}. We further define the sky coverage as the ratio
between the number of pixels with at least 2 observations (which can come either from different passes of the drift scan or different time dumps of the same pass) 
and the total number of pixels in the sky.

In the lower right panel of Fig. \ref{fig:forecast} we show the distribution
over frequencies of the total sky coverage, median noise, and median
signal-to-noise for our forecasts simulations.
We find an overall median signal-to-noise of 2.3. The median signal-to-noise 
is 6.5 at the Galactic plane, 2.8 -- 3.4 at 20 degrees, 2.0 -- 2.6 at 40 degees
and 1.6 -- 2.5 at 60 degrees off the plane. The best frequencies in the ranges
990 MHz to 1010 MHz and 1185 MHz to 1195 MHz achieve a median signal-to-noise of approximately 9 and 8, respectively.
For the noise in our maps, we find a median error of 0.24 K and a median
effective integration time of 149 sec. The best frequency range from 1185 MHz
to 1195 MHz reaches an error of 0.09 K.
Finally, the median sky coverage is given by 57\% and the highest values of 70\% are achieved by the channels 
from 990 MHz to 1010 MHz.

We also study systematic differences between the forecasted and the map without observational effects by 
looking at the difference map between them shown in the lower left panel of Fig. \ref{fig:forecast}. 
We find that the systematic differences in the maps are due to cumulated inaccuracies of the reconstruction 
of baseline and gain described in section \ref{sec:simulatingbgs_seek} as well as unmitigated RFI. The mild 
elevation dependent stripes in Fig. \ref{fig:forecast} imply that the dominant contribution to the systematics is 
coming from the baseline correction which is the only elevation-dependent artifact in the simulations. 
We defer the development of a more complicated and realistic baseline removal to
future work. This uncertainty associated with the baseline removal increases
the total noise and lowers the signal-to-noise ratio of the maps.
The overall median signal-to-noise with systematic uncertainty is 2.0, with a
median signal-to-noise of 6.3 at the Galactic plane, 2.8 -- 3.5 at 20 degrees,
2.0 -- 2.7 at 40 degees and 1.7 -- 2.6 at 60 degrees off the plane. A median
signal-to-noise of 5 -- 6 is achieved in the cleanest frequency channels.

\section{Conclusion}
\label{sec:conclusion}

In this paper, we present two software packages, \hide \& \seek, for simulating and processing data 
from single-dish radio surveys. \hide simulates the entire system chain of a radio telescope from 
the astronomical signal to the time-ordered-data (TOD). \seek on the other hand processes simulated 
and real observed TOD to \healpix maps while calibrating the signal and automatically masking 
contamination from radio frequency interference (RFI). The two packages together provide an 
end-to-end forward-modeling framework that can be used to systematically understand and test the 
various steps in the data processing procedure. They can also be used independently if only simulation 
or data processing is needed. The packages were developed based on the data taken at the Bleien 
Observatory, but can be easily adjusted to model and process similar projects.
The other strength of using both \hide \& \seek is the possibility to perform cross-validation between the 
simulation and the analysis algorithm. We demonstrate how the two packages are used together to study 
systematic effects such as imperfect baseline removal and RFI leakage. 

We present the main architecture of the codes and how typical data is simulated and processed. We then 
describe more specifically how we apply the two packages to a forward-modeling exercise of a Galactic 
survey at 990 -- 1260 MHz. We use \hide to simulate the entire survey with the main survey characteristics matched to 
the Science Verification data from the Bleien Observatory. We then process this simulated data with \seek, 
using settings close to that 
used for the observed data. The result of this forward model is a forecast of the expected output from an 
idealized Galactic survey, with simplifications that are well-understood. We predict a median sky coverage of 50\% 
and a median signal-to-noise ratio of 2.3. In the channels from 990 MHz to 1010 MHz we expect a total 
sky coverage up to 70\% with median signal-to-noise ratio of approximately 9 without systematic errors and 5 -- 6 with systematic errors. 

\hide \& \seek follow common software engineering best practices. Being compliant with well-established 
coding standards, they offer great flexibility for defining data processing pipelines. Although some of the 
current implementations in \hide \& \seek are relatively simple and more general compared to existing 
software such as \cite{offringa2010lofar, peck2013serpent, mcmullin2007casa}, they have the advantage 
of being open-source code with a rigorous structure, and thus provide an easy-to-use foundation to build 
upon for more complicated functionalities. We have developed both packages in pure Python and increase 
the performance of computationally intensive parts by just-in-time compilation. By doing so, we are able to 
perform the RFI mitigation of the TOD in \seek at an rate of 190-200 GB/h/CPU. Using \ivy's parallelization 
scheme, we can furthermore distribute the workload to multiple cores and make efficient use of the available 
hardware. This allows us to easily process the expected data volume from a
five-month survey on a modern laptop.
In \hide we implemented an efficient beam convolution on the sphere by using \textit{Quaternions} combined 
with a KD-Tree. Creating the simulations for the Galactic survey with \hide took around 3 hours on a 
single-core of an average notebook. Further information such as documentation of
the software can be found at \url{http://hideseek.phys.ethz.ch/}.

\appendix

\section{Implementation of beam convolution on a sphere}
\label{sec:beam_convolution}

In order to convolve the beam response with the simulated astronomical signal, we have to rotate the grid 
that defines the beam geometry on a sphere. As this step is repeated for every telescope pointing defined 
by the scanning strategy as well as for every simulated frequency, high efficiency for the operation is crucial. 

Conventionally, spherical rotations are implemented by using the Euler matrix rotation. Applying this rotation 
scheme to a \healpix map requires transforming the spherical pointing angles ($\theta$ and $\phi$) into Euler 
coordinates ($x,y,z$). This involves computing the rotation matrix $R$, applying this matrix to the coordinate 
vector, transforming the results back into spherical coordinates, and finally performing the convolution. 
Typically, applying rotation on multiple axes requires a repetition of the above steps for each axis. 
This can be computationally even more demanding and numerically less stable. 

We have implemented all the coordinate rotations with \textit{Quaternions}, a technique commonly used in 3D 
computer vision \cite{shoemake1985animating}. Rotations over multiple axes can be concentrated into one 
operation with \textit{Quaternions}, which make them computationally more efficient and stable. Furthermore 
they do not suffer from the so-called gimbal lock as Euler rotation do, where a rotation along one axis may produce 
coordinates that do not allow further rotations \cite{shoemake1985animating}. 

Additionally, this allowed us to easily implement a lookup table with a binary search for the $sine$ and $cosine$ 
function to further speed up the rotations. In order to efficiently find the pixels relevant for the rotation of the 
beam geometry we also use store all the \healpix pixel information in a KD-Tree adapted for spherical coordinates.

\section{RFI masking}
\label{sec:rfi}

The \seek package was designed to be able to process data that is heavily contaminated with RFI. \seek's 
automated RFI masking mechanism is based on two steps: the \textsc{SumThreshold} algorithm and a 
\healpix pixel-based outlier rejection. The \textsc{SumThreshold} method was first proposed in \cite{offringa2010post} 
and since then successfully implemented in different radio processing pipelines 
\citep[e.g.,][]{offringa2010lofar, peck2013serpent}. Its simple concept and high detection efficiency makes it an 
attractive candidate for the \seek package.

\textsc{SumThreshold} gradually builds a mask that flags the unwanted RFI in the data. The underlying assumption 
of the algorithm is that the astronomical signal is relatively smooth in both time and frequency (i.e., on the TOD plane), 
while RFI signals have sharp edges. Under this assumption, \textsc{SumThreshold} iteratively improves a model of 
the true astronomical signal by smoothing the TOD with a Gaussian filter. It then clips pixels that lie above a certain 
threshold after subtracting this model from the data. The masking naturally starts with localized, strong RFI bursts. 
The threshold is progressively lowered in each iteration such that more RFI is detected and masked along the process. 
We also take into account that RFI often extends in both time and frequency directions. That is, the algorithm combines 
neighboring pixels and masks them if their sum exceeds a threshold. Due to this underlying assumption \textsc{SumThreshold} has to be applied with
precaution to spectral line data, which can show non-smooth signatures. RFI
originating from radio point sources, such as planes or satellites is well
captured. However, depending on the instrument configuration natural point
sources could be incorrectly masked by the algorithm.

Finally, we apply a
morphological dilation of the mask, i.e.
we enlarge the mask depending on the density of masked pixels in a region in order to capture possible missed RFI leakage. The iterative Gaussian smoothing in the presence of 
a mask is the most computationally intensive part. In order to accelerate the process while keeping the code fully Python, 
we use the \hope just-in-time compiler package \citep{akeret2015hope}, which translates the Python code into C++ and 
compiles at runtime. This allows \seek to perform the RFI mitigation step at a rate of 190-200 GB/h/CPU, which is 
faster than other existing Python implementations of the similar algorithm
\cite[e.g.][]{peck2013serpent}. 

The second step in the RFI mitigation is applied during the \healpix map-making process. Pixel values $X_{0..n}$ 
belonging to the same \healpix pixel and same frequency are analyzed in order to remove outliers. We use a simple but robust outlier 
criterion defined through $\omega_i$, where
\begin{equation*}
  \omega_i=\begin{cases}
    True, & \text{if $X_{i}' / \hat{X'}>t$}\\
    False, & \text{otherwise}
  \end{cases},
\end{equation*}
with
\begin{equation*}
	X'_i = X_i - \hat{X},
\end{equation*}
where $\hat{X}$ is the median of $X_{0..n}$, $\hat{X'}$ the median of $X_{0..n}'$ and $t$ a predefined 
threshold. This replaces the mean and standard deviation often used in sigma clipping, which can be biased 
toward the outliers. 

\section{Flux calibration}
\label{sec:flux_calibration}

We are interested in converting the raw data value $M$ that is recorded in the native instrument units of 
ADU into a brightness temperature $T_{s}$ with the units of Kelvin -- this process is referred as flux calibration. 
Brightness temperature is a measure of the luminosity of the source under a given resolution, and is independent 
of other instrument properties. We refer to this conversion as ``Gain'' $G$, or
\begin{equation}
T_{s}(\lambda) = G(\lambda) M(\lambda) 
\end{equation}

For the data set used in this work, calibration data taken during the survey consists of several transit 
measurements of 
astronomical sources with well known spectra. Transit measurements are taken with the telescope 
parked at some position on the sky where the source of interest is expected to cross due to the rotation 
of the Earth. The data from these transits give an estimate of the beam response profile. Since we know 
the spectra of these sources, we can derive $G$ given our knowledge of the beam. 
\begin{equation}
G(\lambda) = \frac{A_{e}(\lambda)}{2k} \frac{F_{ref}(\lambda)}{M_{ref}(\lambda)}10^{-26}
\end{equation}
where $\lambda$ is the wavelength in meters, $A_{e}$ is the effective aperture size of the instrument 
in square meters, $k$ is the Planck constant, $F_{ref}$ is the spectral density of the reference source 
in Jansky (10$^{-26} W/m^{2}/Hz$), and $M_{ref}$ is the measured ADU value at the peak of the transit 
for the reference source. $A_{e}$ is defined through
\begin{equation}
A_{e}(\lambda) = \frac{\lambda^{2}}{\Omega_{A}},
\end{equation}
where $\Omega_{A}$ is the beam solid angle defined
\begin{equation}
\Omega_{A} = \int\int_{2 \pi} P_{n}(\Omega) d \Omega,
\end{equation}
and $P_{n}$ is the beam profile with its peak normalized to 1. 

We calculate $A_{e}$ assuming the an Airy disk profile with size measured from transit data of the 
Sun. For $M_{ref}$, since we are only interested in the main beam here, we fit the transit data with a 
Gaussian profile. 

\section{Apply \seek to SV data at the Bleien Observatory}
\label{sec:comparison}

\begin{figure}[t]
\begin{center}
	\includegraphics[width=0.5\linewidth]{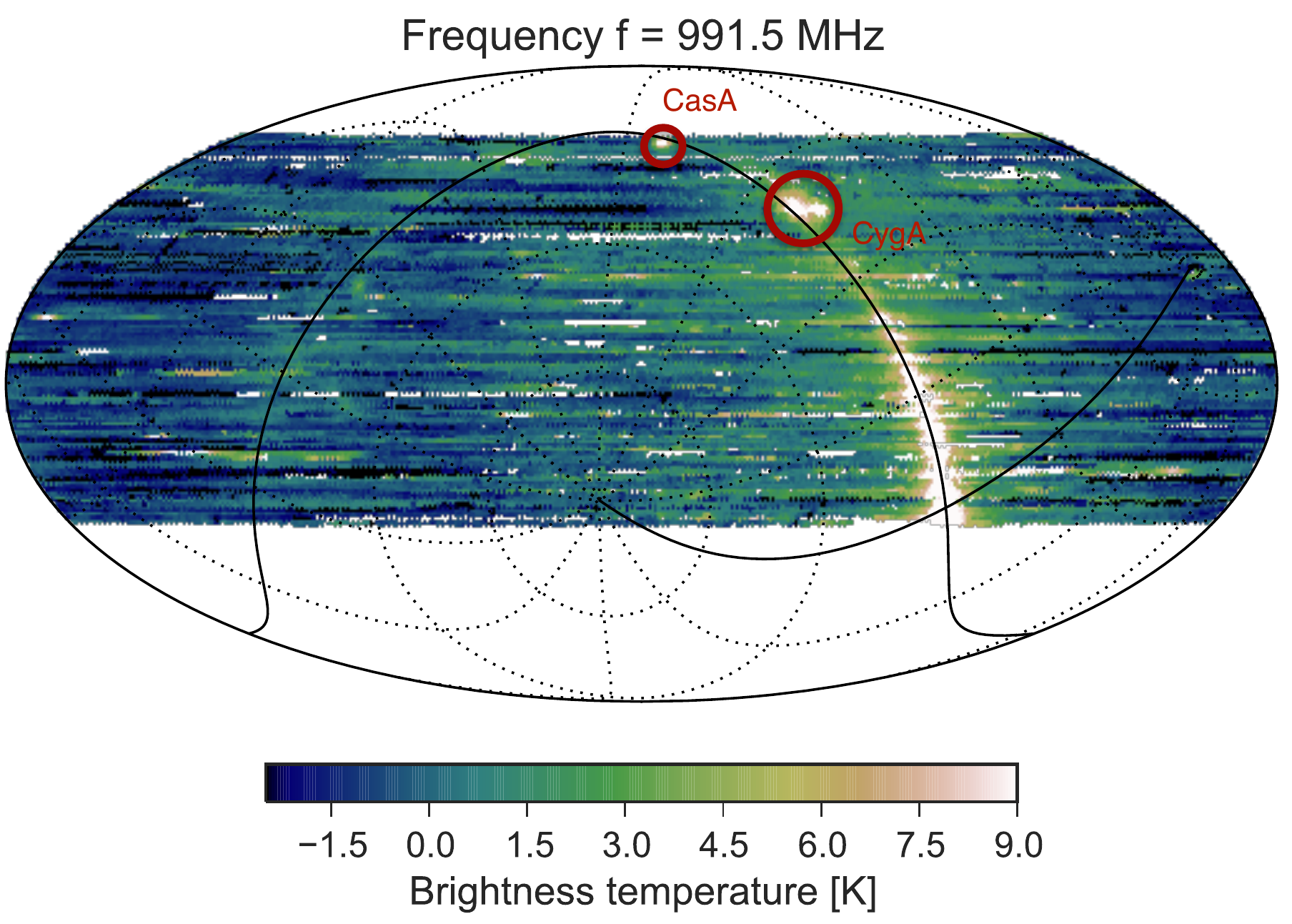}
\end{center}
\caption{Map from the SV data at $\sim 991.5$ MHz.}
\label{fig:bgs_sv}
\end{figure}

We have used \seek to process the SV data from Bleien (see C16 for details of the SV data) to 
demonstrate the usability of \seek on real data.  
Fig. \ref{fig:bgs_sv} shows the resulting map at $\sim 991.5$ MHz with $\sim1$MHz bandwidth  
at a \healpix resolution of \texttt{nside}$=64$. The Galactic plane is clearly visible, as well as the brightest point 
sources close to the plane such as Cassiopeia A and Cygnus A. There are, however, clear patterns of stripes 
in the declination direction in the map, which corresponds to the drift-scan pattern used in the SV data. These 
patterns originate from a time-varying baseline which varies on the time scale of several hours and cannot 
be removed with our simple baseline-removal method in \seek. The baseline also appears to be correlated 
with the ambient temperature and thus higher during day time. This coincides with the time where the 
Galactic center is observed in the SV months, making the Galactic center brighter than expected. 
Furthermore, comparing Fig. \ref{fig:bgs_sv} and Fig. \ref{fig:forecast}, we see a higher masked area in 
the data, which originates from artifacts in the data that were manually masked. These artifacts appear 
as broad-band, bright bursts that are potentially related to the moisture on the cover of the feed horn. 
  
After a quantitative analysis of the maps from SV, we find that the 
statistical uncertainty of the maps are close to expectation from the simulations. However, the 
systematic uncertainties associated with the varying baseline will limit the usage of these maps in 
cosmological studies. We conclude that the stability of the overall system needs to be improved. Although 
the electronic chain described in C16 is stable, the existing observatory 
environment and other supporting infrastructure still need further upgrades to meet the stringent 
requirements set for cosmological surveys.

Below, we discuss several aspects of the system stability that we plan on improving in 
the future. In parallel, we will also work on incorporating these effects in the \hide simulations and 
developing more sophisticated baseline-removal techniques in \seek.  
First, the temperature of the main electronics is only controlled by a moderate heating system 
which can regulate temperature swings up to 40 degrees. Unfortunately, during the survey, there were 
several occasions where the temperature change exceeded this range and the electronic temperature 
was no longer regulated, which in turn resulted in fluctuations of the baseline. Furthermore, temperature 
at the feed horn is not controlled, which adds to the baseline instability. 
Second, the frontend electronics were not properly shielded. This means that water from rain and dew 
leaks in under extreme weather conditions, and the frontend is more heated by the Sun at certain pointing 
directions than others. During the SV observations, there were incidents of heavy rainfall that caused the 
frontend to response strangely and/or stop responding.    

All these valuable lessons learned from studying the SV data will feed into the next stage of 
observations at the Bleien Observatory as well as other future surveys.

\section{Code distribution}
\label{sec:distribution}

Detailed documentation and installation instructions can be found on the package website 
\url{http://hideseek.phys.ethz.ch/}. 
The \hide \& \seek packages are released under the GPLv3 license The 
development is coordinated on GitHub \url{http://github.com/cosmo-ethz/hide} and 
\url{http://github.com/cosmo-ethz/seek}. We welcome all contributions to the code development.

\section{TOD file format}
\label{sec:fileformat}

The default TOD file format and structure used in \hide \& \seek are designed to be consistent with 
that collected at the Bleien Observatory. All files are in HDF5 file formats, and are compressed using the 
common lossless gzip compression. Each HDF5 file contains several ``dataset'' units. Four datasets 
are used to store the time-frequency planes (``P/Phase0'', ``P/Phase1'', `P2/Phase0'', and ``P2/Phase1'') 
and an additional dataset for the time-axis that records the time-stamp for each pixel (``TIME'').
The structure of the four time-frequency planes originated from the special design of the BGS 
project\footnote{These four time-frequency planes include both the measured spectrum ($P$) and the 
measured squared spectrum ($P^{2}$) for both voltage phases ($Phase0$ and $Phase1$).}, 
but one can easily adapt \hide \& \seek to a different file format.

\vspace{0.2in}

\bibliographystyle{elsarticle-num}
\bibliography{hide_seek}

\begin{thebibliography}{10}
\expandafter\ifx\csname url\endcsname\relax
  \def\url#1{\texttt{#1}}\fi
\expandafter\ifx\csname urlprefix\endcsname\relax\def\urlprefix{URL }\fi
\expandafter\ifx\csname href\endcsname\relax
  \def\href#1#2{#2} \def\path#1{#1}\fi

\bibitem{reinecke2006simulation}
M.~Reinecke, K.~Dolag, R.~Hell, M.~Bartelmann, T.~En{\ss}lin, A simulation
  pipeline for the planck mission, Astronomy \& Astrophysics 445~(1) (2006)
  373--373.

\bibitem{nord2016spokes}
B.~Nord, A.~Amara, A.~R{\'e}fr{\'e}gier, L.~Gamper, L.~Gamper, B.~Hambrecht,
  C.~Chang, J.~Forero-Romero, S.~Serrano, C.~Cunha, et~al., Spokes: An
  end-to-end simulation facility for spectroscopic cosmological surveys,
  Astronomy and Computing 15 (2016) 1--15.

\bibitem{bridle2009handbook}
S.~Bridle, J.~Shawe-Taylor, A.~Amara, D.~Applegate, S.~T. Balan, J.~Berge,
  G.~Bernstein, H.~Dahle, T.~Erben, M.~Gill, et~al., Handbook for the great08
  challenge: An image analysis competition for cosmological lensing, The Annals
  of Applied Statistics (2009) 6--37.

\bibitem{refregier2014way}
A.~Refregier, A.~Amara, A way forward for cosmic shear: Monte-carlo control
  loops, Physics of the Dark Universe 3 (2014) 1--3.

\bibitem{bruderer2015calibrated}
C.~Bruderer, C.~Chang, A.~Refregier, A.~Amara, J.~Berge, L.~Gamper, Calibrated
  ultra fast image simulations for the dark energy survey, arXiv preprint
  arXiv:1504.02778.

\bibitem{peterson2015simulation}
J.~R. {Peterson}, J.~G. {Jernigan}, S.~M. {Kahn}, A.~P. {Rasmussen}, E.~{Peng},
  Z.~{Ahmad}, J.~{Bankert}, C.~{Chang}, C.~{Claver}, D.~K. {Gilmore},
  E.~{Grace}, M.~{Hannel}, M.~{Hodge}, S.~{Lorenz}, A.~{Lupu}, A.~{Meert},
  S.~{Nagarajan}, N.~{Todd}, A.~{Winans}, M.~{Young}, {Simulation of
  Astronomical Images from Optical Survey Telescopes Using a Comprehensive
  Photon Monte Carlo Approach}, \apjs 218 (2015) 14.
\newblock \href {http://arxiv.org/abs/1504.06570} {\path{arXiv:1504.06570}},
  \href {http://dx.doi.org/10.1088/0067-0049/218/1/14}
  {\path{doi:10.1088/0067-0049/218/1/14}}.

\bibitem{Battye2013}
R.~A. {Battye}, I.~W.~A. {Browne}, C.~{Dickinson}, G.~{Heron}, B.~{Maffei},
  A.~{Pourtsidou}, {H I intensity mapping: a single dish approach}, \mnras 434
  (2013) 1239--1256.
\newblock \href {http://arxiv.org/abs/1209.0343} {\path{arXiv:1209.0343}},
  \href {http://dx.doi.org/10.1093/mnras/stt1082}
  {\path{doi:10.1093/mnras/stt1082}}.

\bibitem{Santos2015}
M.~{Santos}, P.~{Bull}, D.~{Alonso}, S.~{Camera}, P.~{Ferreira}, G.~{Bernardi},
  R.~{Maartens}, M.~{Viel}, F.~{Villaescusa-Navarro}, F.~B. {Abdalla},
  M.~{Jarvis}, R.~B. {Metcalf}, A.~{Pourtsidou}, L.~{Wolz}, {Cosmology from a
  SKA HI intensity mapping survey}, Advancing Astrophysics with the Square
  Kilometre Array (AASKA14) (2015) 19\href {http://arxiv.org/abs/1501.03989}
  {\path{arXiv:1501.03989}}.

\bibitem{Bigot-Sazy2016}
M.-A. {Bigot-Sazy}, Y.-Z. {Ma}, R.~A. {Battye}, I.~W.~A. {Browne}, T.~{Chen},
  C.~{Dickinson}, S.~{Harper}, B.~{Maffei}, L.~C. {Olivari}, P.~N.
  {Wilkinsondagger}, {HI Intensity Mapping with FAST}, in: L.~{Qain}, D.~{Li}
  (Eds.), Frontiers in Radio Astronomy and FAST Early Sciences Symposium 2015,
  Vol. 502 of Astronomical Society of the Pacific Conference Series, 2016,
  p.~41.
\newblock \href {http://arxiv.org/abs/1511.03006} {\path{arXiv:1511.03006}}.

\bibitem{swinbank2015lofar}
J.~D. {Swinbank}, T.~D. {Staley}, G.~J. {Molenaar}, E.~{Rol}, A.~{Rowlinson},
  B.~{Scheers}, H.~{Spreeuw}, M.~E. {Bell}, J.~W. {Broderick}, D.~{Carbone},
  H.~{Garsden}, A.~J. {van der Horst}, C.~J. {Law}, M.~{Wise}, R.~P. {Breton},
  Y.~{Cendes}, S.~{Corbel}, J.~{Eisl{\"o}ffel}, H.~{Falcke}, R.~{Fender}, J.-M.
  {Grie{\ss}meier}, J.~W.~T. {Hessels}, B.~W. {Stappers}, A.~J. {Stewart},
  R.~A.~M.~J. {Wijers}, R.~{Wijnands}, P.~{Zarka}, {The LOFAR Transients
  Pipeline}, Astronomy and Computing 11 (2015) 25--48.
\newblock \href {http://arxiv.org/abs/1503.01526} {\path{arXiv:1503.01526}},
  \href {http://dx.doi.org/10.1016/j.ascom.2015.03.002}
  {\path{doi:10.1016/j.ascom.2015.03.002}}.

\bibitem{mcmullin2007casa}
J.~P. {McMullin}, B.~{Waters}, D.~{Schiebel}, W.~{Young}, K.~{Golap}, {CASA
  Architecture and Applications}, in: R.~A. {Shaw}, F.~{Hill}, D.~J. {Bell}
  (Eds.), Astronomical Data Analysis Software and Systems XVI, Vol. 376 of
  Astronomical Society of the Pacific Conference Series, 2007, p. 127.

\bibitem{dodson2016imaging}
R.~{Dodson}, K.~{Vinsen}, C.~{Wu}, A.~{Popping}, M.~{Meyer}, A.~{Wicenec},
  P.~{Quinn}, J.~{van Gorkom}, E.~{Momjian}, {Imaging SKA-scale data in three
  different computing environments}, Astronomy and Computing 14 (2016) 8--22.
\newblock \href {http://arxiv.org/abs/1511.00401} {\path{arXiv:1511.00401}},
  \href {http://dx.doi.org/10.1016/j.ascom.2015.10.007}
  {\path{doi:10.1016/j.ascom.2015.10.007}}.

\bibitem{chang2016an}
C.~{Chang}, C.~{Monstein}, J.~{Akeret}, S.~{Seehars}, A.~{Refregier},
  A.~{Amara}, A.~{Glauser}, B.~{Stuber}, {An Integrated System at the Bleien
  Observatory for Mapping the Galaxy}, ArXiv e-prints\href
  {http://arxiv.org/abs/1607.07451} {\path{arXiv:1607.07451}}.

\bibitem{Gorski2005}
K.~M. {G{\'o}rski}, E.~{Hivon}, A.~J. {Banday}, B.~D. {Wandelt}, F.~K.
  {Hansen}, M.~{Reinecke}, M.~{Bartelmann}, {HEALPix: A Framework for
  High-Resolution Discretization and Fast Analysis of Data Distributed on the
  Sphere}, \apj 622 (2005) 759--771.
\newblock \href {http://arxiv.org/abs/astro-ph/0409513}
  {\path{arXiv:astro-ph/0409513}}, \href {http://dx.doi.org/10.1086/427976}
  {\path{doi:10.1086/427976}}.

\bibitem{Airy1838}
G.~B. {Airy}, {Ueber die Diffraction eines Objectivs mit kreisrunder Apertur},
  Annalen der Physik 121 (1838) 86--95.
\newblock \href {http://dx.doi.org/10.1002/andp.18381210906}
  {\path{doi:10.1002/andp.18381210906}}.

\bibitem{deOliveira-Costa2008}
A.~{de Oliveira-Costa}, M.~{Tegmark}, B.~M. {Gaensler}, J.~{Jonas}, T.~L.
  {Landecker}, P.~{Reich}, {A model of diffuse Galactic radio emission from 10
  MHz to 100 GHz}, \mnras 388 (2008) 247--260.
\newblock \href {http://arxiv.org/abs/0802.1525} {\path{arXiv:0802.1525}},
  \href {http://dx.doi.org/10.1111/j.1365-2966.2008.13376.x}
  {\path{doi:10.1111/j.1365-2966.2008.13376.x}}.

\bibitem{Wells1981}
D.~C. {Wells}, E.~W. {Greisen}, R.~H. {Harten}, {FITS - a Flexible Image
  Transport System}, \aaps 44 (1981) 363.

\bibitem{offringa2010post}
A.~Offringa, A.~de~Bruyn, M.~Biehl, S.~Zaroubi, G.~Bernardi, V.~Pandey,
  Post-correlation radio frequency interference classification methods, Monthly
  Notices of the Royal Astronomical Society 405~(1) (2010) 155--167.

\bibitem{offringa2010lofar}
A.~Offringa, A.~de~Bruyn, S.~Zaroubi, M.~Biehl, A lofar rfi detection pipeline
  and its first results, arXiv preprint arXiv:1007.2089.

\bibitem{peck2013serpent}
L.~W. Peck, D.~M. Fenech, Serpent: Automated reduction and rfi-mitigation
  software for e-merlin, Astronomy and Computing 2 (2013) 54--66.

\bibitem{shoemake1985animating}
K.~Shoemake, Animating rotation with quaternion curves, in: ACM SIGGRAPH
  computer graphics, Vol.~19, ACM, 1985, pp. 245--254.

\bibitem{akeret2015hope}
J.~Akeret, L.~Gamper, A.~Amara, A.~Refregier, Hope: A python just-in-time
  compiler for astrophysical computations, Astronomy and Computing 10 (2015)
  1--8.

\end{thebibliography}

\end{document}